\documentclass{aa}
\bibpunct{(}{)}{;}{a}{}{,} 

\usepackage{graphicx}
\usepackage[varg]{txfonts}
\def\kms{km$\,$s\ensuremath{^{-1}}}

\def\Ha{${\rm H}{\alpha}$}
\def\Hb{${\rm H}{\beta}$}

\def\Mgg{{\rm Mg}}
\def\Mgb{{\rm Mg}$b$}
\def\Mg2{{\rm Mg}$_2$}

\def\oiiipg{[O~{\small III}]$\,\lambda\lambda4959,5007$}

\def\oiiig{[O~{\small III}]$\,\lambda5007$}
\def\oiii{[O~{\small III}]}
\def\niipg{[N~{\small II}]$\,\lambda\lambda5198,5200$}
\def\oipg{[O~{\small I}]$\,\lambda\lambda6300,6363$}
\def\niiipg{[N~{\small II}]$\,\lambda\lambda6548,6583$}
\def\niiip{[N~{\small II}]$\,\lambda6548$}
\def\niiig{[N~{\small II}]$\,\lambda6583$}
\def\nii{[N~{\small II}]}
\def\siipg{[S~{\small II}]$\,\lambda\lambda6716,6731$}
\def\siip{[S~{\small II}]$\,\lambda6716$}

\begin{document}

\title{Evidence for the formation of the young counter-rotating
  stellar disk from gas acquired by IC 719\thanks{Based on observation
    collected at the European Southern Observatory for the programme
    095.B-0686(A) } }

   \author{A. Pizzella\inst{1,2} 
        \and
        L. Morelli\inst{1,2}
        \and
        L. Coccato\inst{3}
        \and 
         E. M. Corsini\inst{1,2}
        \and
         E. Dalla Bont\`a\inst{1,2}
        \and
        M. Fabricius\inst{4,5}
        \and
        R. P. Saglia\inst{4,5}
        }

  \institute{Dipartimento di Fisica e Astronomia ``G. Galilei'',
    Universit\`a di Padova, vicolo dell'Osservatorio 3, I-35122
    Padova, Italy\\
    \email{alessandro.pizzella@unipd.it}
    \and INAF-Osservatorio Astronomico di Padova, vicolo dell'Osservatorio 5, I-35122 Padova, Italy
    \and European Southern Observatory, Karl-Schwarzschild-Strasse 2, 85748 Garching, Germany
    \and Max-Planck-Insitut fur extraterrestrische Physik, Giessenbachstrasse, D-85748 Garching, Germany
    \and Universit\"ats-Sternwarte M\"unchen, Scheinerstrasse 1, D-81679 M\"unchen, Germany
     }

   \date{Received ...; accepted ...}

 
  \abstract
  {}
  {The formation scenario of extended counter-rotating stellar disks
    in galaxies is still debated. In this paper, we study the S0
    galaxy IC~719 known to host two large-scale counter-rotating
    stellar disks in order to investigate their formation mechanism.}
  {We exploit the large field of view
  and wavelength coverage of the Multi Unit Spectroscopic Explorer
  (MUSE) spectrograph to derive two-dimensional (2D) maps of the various
  properties of the counter-rotating stellar disks, such as age, metallicity,
  kinematics, spatial distribution, the kinematical and chemical
  properties of the ionized gas, and the dust map.}
  {Due
  to the large wavelength range, and in particular to the presence of
  the Calcium Triplet $\lambda\lambda8498, 8542, 8662\,$\AA\ (CaT hereafter), the
  spectroscopic analysis allows us to separate the two stellar components
  in great detail. This permits precise measurement of both the velocity
  and velocity dispersion of the two components as well as their spatial distribution.
  We derived a 2D map of the age and metallicity of the two stellar
  components as well as the star formation rate and gas-phase metallicity from the
  ionized gas emission maps.}
  {The main stellar disk of the galaxy is kinematically hotter, older,
    thicker and with larger scale-length than the secondary disk. There is no doubt
    that the
    latter is strongly linked to the ionized gas component: they have the same
    kinematics and similar vertical and radial spatial distribution.  This result is in
    favor of a gas accretion scenario over a binary merger scenario to
    explain the origin of counter-rotation in IC~719. One     source of gas that may have contributed to the accretion process is the
    cloud that surrounds IC~719.}

 

  \authorrunning{A. Pizzella et al.}

   \maketitle
%

\section{Introduction}
\label{sec:intro}
It is widely accepted that galaxies undergo a number of acquisition
events during their evolution and growth.  Depending on the nature of
the accreted material and on the geometry of the process, such events
may leave temporary or permanent signatures on the morphology,
kinematics, and stellar population properties of the host
galaxy. Among these, the presence of structural components with
  remarkably different kinematics represents a fascinating case. A
  large variety of phenomena enters in this category, such as
  counter-rotating gas disks \citep[][]{1995Natur.375..661C,
    2012MNRAS.422.1083C}, counter-rotating stellar disks
  \citep{1992ApJ...394L...9R, 1996ApJ...458L..67B}, orthogonal bulges
  \citep{1999ApJ...519L.127B, 2001Ap&SS.276..467S, 2012MNRAS.423L..79C}, orthogonal gaseous
  structures \citep{2002A&A...382..488C} and decoupled stellar cores
  \citep{2000A&A...360..439S}; see \citet{Galletta96,
    Bertola+99} and \citet{Corsini14} for reviews.

  In this paper, we investigate the case of extended
  counter-rotation of large stellar disks. Although some attempts
have been made to explain stellar counter-rotation as due to
self-induced phenomena such as dissolving bars
\citep[e.g.,][]{1994ApJ...420L..67E}, this phenomenon is usually
considered as the signature of a past merging event or accretion
  (e.g., \citealt{1997ApJ...479..702T, 2014MNRAS.437.3596A, 2017MNRAS.471.1892B}).  However, the
  details of the formation mechanisms of large counter-rotating
  stellar disks are still under debate, and their existence still
  represents a puzzle in the broad context of formation and evolution
  of galaxies.
\begin{figure}
\includegraphics[width=9.0cm,angle=0]{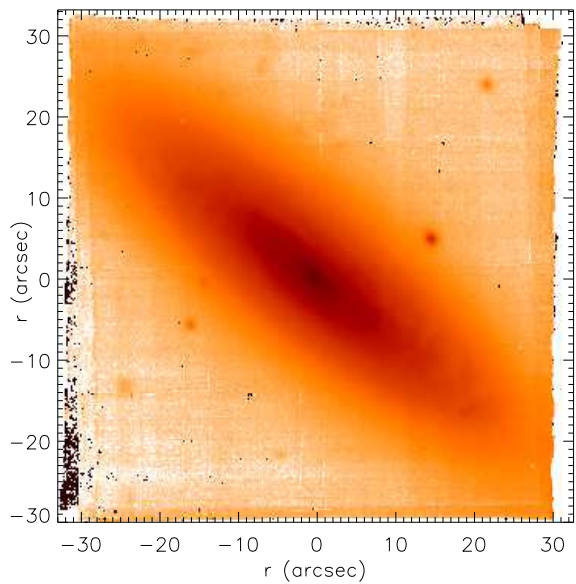}
\caption{MUSE image of IC 719. The image was obtained by
  averaging the wavelength range $7500$---$8000\,$\AA . The
  image intensity is shown with a logarithmic scale. The field of view
  and orientation of all 2D maps of the paper are $\sim
  1'\times 1'$ and North up, East left.}
\label{fig:ic719}
\end{figure}

In the current theoretical picture, two scenarios have been proposed
to explain their formation, each leaving different signatures in the
kinematics and stellar populations: i) binary galaxy major
 mergers (e.g., \citealt{2001Ap&SS.276..909P}); and ii) gas
accretion or gas-rich minor mergers on a pre-existing galaxy followed
by star formation (e.g.,  \citealt{1998ApJ...506...93T,
  2017MNRAS.471.1892B}).

A major merger of two disk galaxies may end up with the formation of a
disk galaxy with a massive counter-rotating disk under certain
conditions.  \citet{2009MNRAS.393.1255C} built a numerical simulation
to explain the case of NGC~4550. They found that two gas-rich disk
galaxies of similar mass can merge and, depending on the geometry of the
encounter, can form a massive disk galaxy with two counter-rotating
stellar disks embedded one into the other. Concerning the fate of the
gas component, the more prominent one, after dissipating with its
counter-rotating counterpart, settles onto the equatorial plane. The
two resulting disks have different thicknesses, the one associated to
the gas being dynamically hotter than the other. In this scenario, the
stellar population properties of the two stellar components depend on
the properties of the progenitors and on star-formation episodes triggered by the merger itself. On the other
  hand, this result holds for this specific case of a perfectly
  co-planar merger, and other possibilities should be investigated
  too.

When a gas-poor disk galaxy acquires gas from the surrounding
environment (either in a short or prolonged episode) or from a
gas-rich minor merger, a counter-rotating stellar disk may form if the
gas angular momentum is opposite to that of the host galaxy. The
newly-formed counter-rotating stellar disk is thinner and dynamically
colder than the prograde disk. The spatial distribution of the
counter-rotating component depends on the initial angular momentum of
the acquired gas and on the amount of gas in the primordial galaxy. In
this scenario, the stellar component associated to the gas disk is
predicted to be the youngest. Moreover, if the secondary stellar disk
has formed recently, it is expected to have similar metallicity to the
gas from which it originated. The analysis of the metallicity of both
the stellar and gas-phase components can be used to identify the gas
reservoir that formed the prograde and/or retrograde stellar
components \citep{2013A&A...549A...3C}. \citet{2014MNRAS.437.3596A}
considered the case where gas is acquired in a single episode and found
that it may give rise to two counter-rotating stellar disks with
different scale lengths and significantly different ages.  A similar
result is found by \citet{2017MNRAS.471.1892B} simulating a 1:10 minor
merger between a gas-poor galaxy and a less-massive gas-rich one.

One key aspect in this investigation is to study the
  counter-rotating components independently by removing their mutual
  contamination on the total observed spectrum along the
  line of sight (LOS).  Indeed, in recent years we have developed a
spectroscopic decomposition technique that fits the kinematics and the
stellar populations of two stellar components, and therefore separates
their contribution from the total observed spectrum
\citep{2011MNRAS.412L.113C}. The spectroscopic decomposition
  technique is particularly suitable in the study of extended
counter-rotating disks where the LOS velocity of the two
stellar components and their low velocity dispersion allows a reliable
spectroscopic separation. It has been further developed and
  successfully applied by several teams to a number of disk galaxies
  with stellar counter-rotation: NGC 524 \citep{Katkov+11}, IC~719 \citep{2013ApJ...769..105K},
  NGC~448 \citep{2016MNRAS.461.2068K} NGC~1366
\citep{2017A&A...600A..76M}, NGC~3593 \citep{2013A&A...549A...3C},
NGC~4138  \citep{2014A&A...570A..79P}, NGC~4191
\citep{2015A&A...581A..65C}, NGC~4550 \citep{2013A&A...549A...3C,
  2013MNRAS.428.1296J},  NGC 5102 \citep{Mitzkus+17}, and
NGC~5719 \citep{2011MNRAS.412L.113C}.
The technique was also applied to study orthogonally decoupled
  structures such as polar rings \citep{Coccato+14} and
  photometrically distinct components such as bulge and disk
  \citep{Fabricius+14, Tabor+17}.

  A lot of importance was given to the age
  and metallicity of the two stellar components, and to the gas-phase
  metallicity in order to identify the most probable formation
  scenario.  In most cases, a significant difference in the age
  of the stellar population of the two components was found.  The
  younger stellar component was always found to be co-rotating with the
  gaseous disk, and  to be the least luminous and least massive. This 
  supports the gas accretion scenario, although without totally
  excluding the binary merging scenario.

  Few exceptions are found to this picture. In NGC~4550 and NGC~448,
  the two counter-rotating stellar disks have similar ages. It is
  interesting to note that they are the oldest systems analyzed to date,
   with $T\sim 7\,$Gyr and $T\sim9\,$Gyr, respectively.  The
  most intriguing exception is NGC~1366. In this galaxy, the ionized gas component is not kinematically associated to
  either of the stellar counter-rotating components. The most plausible
  interpretation is that the gas clouds are still in chaotic motion,
  possibly dissipating to a disk at present
  \citep{2017A&A...600A..76M}. The ambiguity of the results on
  NGC~1366 suggests that additional physical properties are needed to
  establish a direct link between the ionized-gas disk and the
  secondary counter-rotating stellar component, in addition to the
  velocity and stellar population properties of the two components.

For this reason, in this paper we take into account other physical
quantities that play a relevant role in the analysis of such complex
systems in addition to the previous analyzed observables. One of these
is the velocity dispersion of the stellar components.
\citet{2011A&A...535A...5Q} show that, if the counter-rotating stellar
disk recently formed from the gaseous disk as a consequence of a
gas-rich minor merger, we expect that such a newly formed disk is
characterized by a low velocity dispersion and a flattening which
should be similar to those of the gaseous disk. 
On the contrary, an older stellar disk should be thicker and have a larger velocity dispersion due
to dynamical heating processes \citep[e.g.,][]{2012MNRAS.423.2726G}.
In this case, the counter-rotating gaseous disk
is the result of a subsequent acquisition event.  This seems to be the
case for explaining the formation of the counter-rotating disk in NGC
1366 \citep{2017A&A...600A..76M}.

Another important constraint to firmly link the stellar and gas disks
is the present-day star formation activity. If the stellar disk is
still currently forming stars, the consequent star formation
should be evident in the spectra. In some of the targets, we did find
active star formation arranged in a ring-like structure of kiloparsec scale
(NGC~3593, NGC~4138, and NGC~5719). The knowledge of its location and
strength gives important clues about the timing of the formation
process as star formation is the process that transforms the
gas into stars.

The recent construction of integral field unit spectrographs
characterized by a large field of view and extended wavelength range
gives the possibility of a step forward in the study of these
objects. In particular the Multi Unit Spectroscopic Explorer
\citep[MUSE;][]{2010SPIE.7735E..08B} operating on the Very Large
Telescope (VLT) allows us to map counter-rotating galaxies with an
unprecedented efficiency in terms of telescope time, due to the
combination of a large field of view and high optical efficiency. In addition,
the spectral resolution that characterizes MUSE in the CaT wavelength
region allows one to clearly separate these absorption lines as soon as
the difference in velocity of the two stellar components is greater
than $\sim 50\,$\kms .  In the most favorable cases, we can measure
not only the velocity but also the velocity dispersion of the two
stellar disks. This allows us to compare the velocity dispersion of
the stellar and gas disk to obtain crucial constraints on their
connections and formation.

In this work we study the early-type disk galaxy IC~719 (Fig.\ref{fig:ic719}).  It is
classified as S0? by \citet[][hereafter RC3]{1991rc3..book.....D}, and
its apparent size and magnitude are $0\farcm65\, \times\, 0\farcm20$
and $B^0_{T}=14.00$ mag (RC3).  The disk inclination is $i=77^{\circ}$  derived from the ring-like structure visible in the near-infrared (NIR) bands as
shown by \citealp{2013ApJ...769..105K}. This structure is very well spatially defined and represents
the best indicator of galaxy inclination. The galaxy is therefore well
spatially resolved and relatively bright, two necessary conditions to
conduct our analysis. In this work we adopt a photometric position
angle in the sky of $PA=53^{\circ}$ (counted from North to East) and a distance to the galaxy of
$D=28.6$ Mpc (from the NASA/IPAC extragalactic database).

The galaxy is known to have an extended counter-rotating stellar
component and was studied by \citet{2013ApJ...769..105K} by means
of long-slit and SAURON integral field spectroscopy. It therefore represents an ideal target to be re-observed with the capabilities of
the MUSE integral field spectrograph. With our new observations, in
particular with the aid of an extended field of view and wavelength
coverage, we are able to improve the accuracy and precision on the
observables and measure with great precision the properties
(kinematics, morphology, and stellar population) of the two stellar
disks, the star formation rate (SFR), and the gas-phase metallicity.

This paper is organized as follows. The observations and data
reduction are presented in Sect. \ref{sec:data}. In
Sect. \ref{sec:analysis} we describe the data analysis. The results
about the measurements of the distribution, kinematics, and chemical
properties of the stars and ionized gas are given in Sect.
\ref{sec:results}. We discuss our results and present our conclusions
in Sect. \ref{sec:Discussion}.

\begin{figure}
\centering
\includegraphics[width=9.0cm,angle=0]{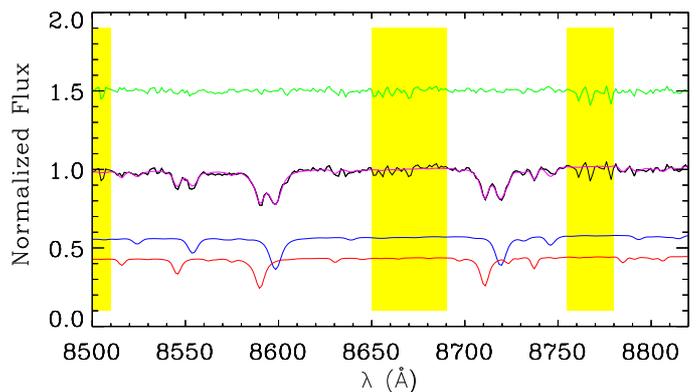}

\caption{Example of the observed (black line) and fitted spectrum in
  a bin where the two stellar components appear separated. Here the
  fit is limited to the CaT wavelength region.  We show the two
  stellar spectra fitted to the data (blue and red lines), and their sum
  (magenta line). The green line represents the
  residuals shifted upward by 1.5 units. The yellow regions are regions
  masked and not included in the fit.}
\label{fig:specfit_CaT}
\end{figure}

\begin{figure*}
\centering
\includegraphics[width=18.0cm,angle=0]{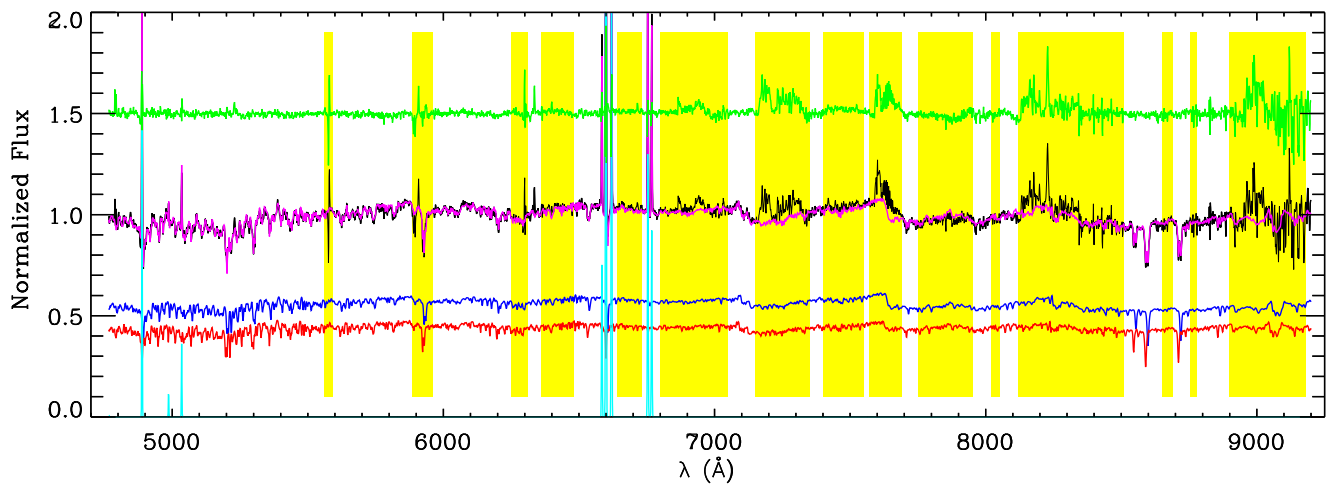}
\caption{As in Fig. \ref{fig:specfit_CaT} but considering the whole
  wavelength extension. The cyan line represents the ionized gas
  emission while the magenta line represents the sum of the two
  stellar components and the ionized gas emission.}
\label{fig:specfit_all}
\end{figure*}
\begin{figure}
\centering
\includegraphics[width=9.0cm,angle=0]{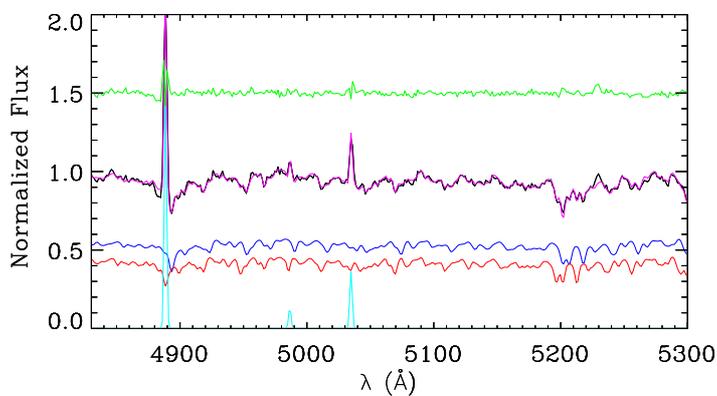}
\caption{Zoom in on the \Hb\ and \Mgg\ spectral region of Fig. \ref{fig:specfit_all}.}
\label{fig:specfit_hb}
\end{figure}

%

\section{Observations and data reduction}
\label{sec:data}

The observations of IC~719 were carried out in service mode at
European Southern Observatory (ESO) in Cerro Paranal with the
UT4-Yepun telescope equipped with MUSE on April 11, 2015.  The
chosen setup was the nominal mode which delivers a wavelength range of
$4800\,$\AA--$9300\,$\AA\ with a resolving power ranging from $1770$ at
$4800\,$\AA\ to $3590$ at $9300\,$\AA.  Spectral sampling is of
1.25\,\AA/pixel while the spatial sampling is
$0\farcs2 \times 0\farcs2$ pixel$^{-2}$. The field of view is
$1\arcmin \times 1\arcmin$. 

The target was acquired in the center of the instrument field of view.
The total integration was split into three different exposures of
840 seconds each. An offset of $1\arcsec$ and a rotation of $90\degr$
was applied between each exposure in order to properly address spatial
instrumental inhomogeneities.  A sky exposure of 300s was taken
during the same observing block with an offset of about $100\arcsec$
with respect to the galaxy center parallel to its minor axis in a sky
region free of bright stars.
 
\subsection{Data reduction}\label{sec:prered}
The raw data from each exposure were pre-reduced and combined with the
MUSE pipeline v1.4 muse/1.1.90 \citep{2012SPIE.8451E..0BW} run under
the EsoReflex environment \citep{2013A&A...559A..96F}.  The reduction
includes bias subtraction, flat-fielding, and illumination correction
(using a twilight exposure), wavelength calibration, and
alignment and a combination of the three exposures.  One standard
star was observed at the beginning of the night
that we used
for flux calibration and correction of telluric features.
The contribution of the night sky has been removed from the spectra by the data
reduction pipeline using the offset sky exposure.

The seeing {\em FWHM\/} read from the ESO seeing monitor was in the
range $1\farcs0 < FWHM < 1\farcs5$ during the observations, consistent
with the value of $FWHM=1\farcs3$ measured on {\it bona-fide}
unresolved H{\small II} regions visible in the final combined
datacube.  We used a sky spectrum, extracted from the datacube before
sky subtraction, to check the spectral resolution.  The spectra of
about 1000 spaxels have been averaged in a region free of galaxy
contamination and we performed a Gaussian fit on the night sky
emission lines. The central wavelength and $FWHM$ of $\sim 140$ lines
were obtained and compared with the line atlas from
\citet{1996PASP..108..277O, 1997PASP..109..614O}. We found mean offset
of $\lambda_{measure}-\lambda_{sky}=-0.40\pm0.09\,$\AA\ constant with
wavelength. The {\em FWHM\/} shows some variation at
  different wavelengths, being $2.65\pm0.08\,$\AA\ at \Hb,
  $2.50\pm0.08\,$\AA\ at \Ha, and $2.39\pm0.08\,$\AA\ in the CaT
  region.  This is in very good agreement with the nominal values of the
  instrument \citep{2010SPIE.7735E..08B}.  The resulting instrumental
  velocity dispersion is $69.4\pm 2.1\,$\kms\ at \Hb, $48.5\pm
  1.5\,$\kms\ at \Ha, and $35.3\pm1.2 \,$\kms\ in the CaT region.

\subsection{Spatial re-sampling of the data cube}\label{sec:red}
After pre-reduction we averaged the one-dimensional (1D) spectra
  of adjacent single spaxels in order to increase the
signal-to-noise ratio $(S/N)$ of the resulting 1D   spectra.  We used the adaptive spatial binning software developed
by \citet{2003MNRAS.342..345C} based on Voronoi tessellation.  We
estimated the signal of each individual spaxel by taking
  the median of the flux in the wavelength range $4762\,$\AA--$6737\,$\AA,
  which is the largest range avoiding the spectral region more
  effected by atmospheric absorption and emission features.

We chose a target $S/N$ of 120 in each bin. As noise we
  adopted the square root of the median of the variance given by the
  pipeline. It must be noted that this is not the S/N of the binned
  spectra. The latter is estimated {\it a posteriori} from the
  residual fitting the galaxy spectra and
  turned out to be $\sim 40$.

We obtained 144 independent spectra. In the center
of the galaxy, where the signal is higher, the area covered by each
bin is about $1$\arcsec$\times 1$\arcsec ($\sim 25$ pixels) while in
more external regions it can be up to about $4$\arcsec$\times
4$\arcsec ($\sim 400$ pixels).
  
\subsection{Optimization of the sky subtraction}\label{sec:sky}
Some residual sky emission was still present at the end of
pre-reduction. We therefore performed a further correction as follows.
We took the average of all the single spectra with absent or very low
signal that for this reason were discarded when building the
spatially binned Voronoi tessellation.  We took this average spectrum
as a template for the ``sky-residual''. We then subtracted this template
from each of the 144 spectra after scaling it with a proper factor
chosen to reduce the scatter in regions free of strong galaxy features
but with a relevant residual of the first sky subtraction above
$7500\,$\AA . The result is a clean spectrum. Spectral regions
affected by relevant sky lines were nonetheless masked when analyzing
the spectra (Sect. \ref{sec:analysis}).
\section{Analysis}
\label{sec:analysis}

\subsection{Stellar and ionized gas kinematics}\label{sec:kinem}
We measured the kinematics of the two stellar components and of the
ionized gas using the implementation of the penalized pixel fitting
code \citep[pPXF,][]{2004PASP..116..138C} developed by
\citet{2011MNRAS.412L.113C}.  In each spatial bin, the code builds the
spectrum of the galaxy as the combination of two synthetic templates,
one for each stellar component. The single template is built as the
linear combination of stellar spectra from the library of
\citet{2005A&A...442.1127M}, characterized by an instrumental $FWHM =
1.35\,$\AA , for the CaT region, and the extended Medium-resolution
INT Library of Empirical Spectra library (MILES) with the canonical
base models {\em BaseFe} \citet{2012MNRAS.424..157V}, with
instrumental $FWHM = 2.54\,$\AA\ \citep{Beifiori+11,
  2011A&A...532A..95F} when considering the whole spectral range.  The
template is convolved with  Gaussian line-of-sight velocity
distributions (LOSVDs).  The LOSVDs used for the two stellar
components have different velocities and velocity dispersions.  Spectra
are normalized to their continuum level at $5100\,$\AA, therefore the
relative contribution of each component to the total spectrum is in
terms of light. Gaussian functions are added to the involved synthetic
templates to account for ionized-gas emission lines and fit
simultaneously to the observed galaxy spectra. Multiplicative Legendre
polynomials are included in the fit for both synthetic templates to
match the shape of the galaxy continuum. The use of multiplicative
polynomials also accounts for the effects of dust extinction and
variations in the instrument transmission.  As a result of the fit,
for each spatial bin, the code returns the spectra of the two best-fit
synthetic stellar templates and ionized-gas emissions, along with
their best-fitting parameters: $Fr$ (that we define as the ratio
between the flux of one component and the total flux), velocity, and
velocity dispersion. Errors on all parameters are computed by the modified pPXF routine
as in \citet{2004PASP..116..138C}.
The line strength of the Lick indices of the two
counter-rotating components are measured from the two best-fit
synthetic templates.  A number of wavelength regions were masked
out when performing the fit because they are affected by poor sky
subtraction. Also, regions with no relevant stellar absorption features
were masked in order to increase the effective $S/N$.  The
procedure was applied in two steps, considering two different
spectral ranges.  The wavelength range $8500\,$\AA--$8800\,$\AA, which
contains prominent Ca absorption-line features, was used to gain
a precise measurement of the kinematics. In fact, the MUSE resolving
power is higher at this wavelength range ($\sigma_{8600\AA} \sim
35\,$\kms) than in the Mg range ($\sigma_{5200\AA} \sim
60\,$\kms). Thus, we better resolve the absorption line features and
obtain more precise measurements of velocity and, in particular,
velocity dispersion of the two stellar components, by measuring the
kinematics in the CaT region. This is why we adopted the
library stellar by \citet{2005A&A...442.1127M} in this spectral
region. In Fig. \ref{fig:specfit_CaT} we plot an example of the
spectral decomposition in a bin positioned along the major axis of the
galaxy where the two components are well separated.  The best fit
velocity and velocity dispersion measured in the CaT region were used
as starting guesses to fit the spectra in the full
$4778\,$\AA--$9200\,$\AA\ wavelength range
(Fig. \ref{fig:specfit_all}). The best fit spectra obtained in this
range were used to measure the ionized-gas kinematics and intensity, and the
luminosity fraction of the two components. In
Fig. \ref{fig:specfit_all} we show the best fit to the full wavelength
range obtained for the same spatial bin as in
Fig. \ref{fig:specfit_CaT}.
To measure the stellar population properties, we repeated the spectral fit
  considering only the $4778\,$\AA--$5450\,$\AA\ wavelength range. In this case, we kept the
  velocity and velocity dispersion fixed to the values measured in the CaT region and
  left all the other parameters free to vary.
In Fig. \ref{fig:specfit_hb} we show the result for the same spatial bin as in
Fig. \ref{fig:specfit_CaT}.

\begin{figure*}
\centering
\includegraphics[width=6.0cm]{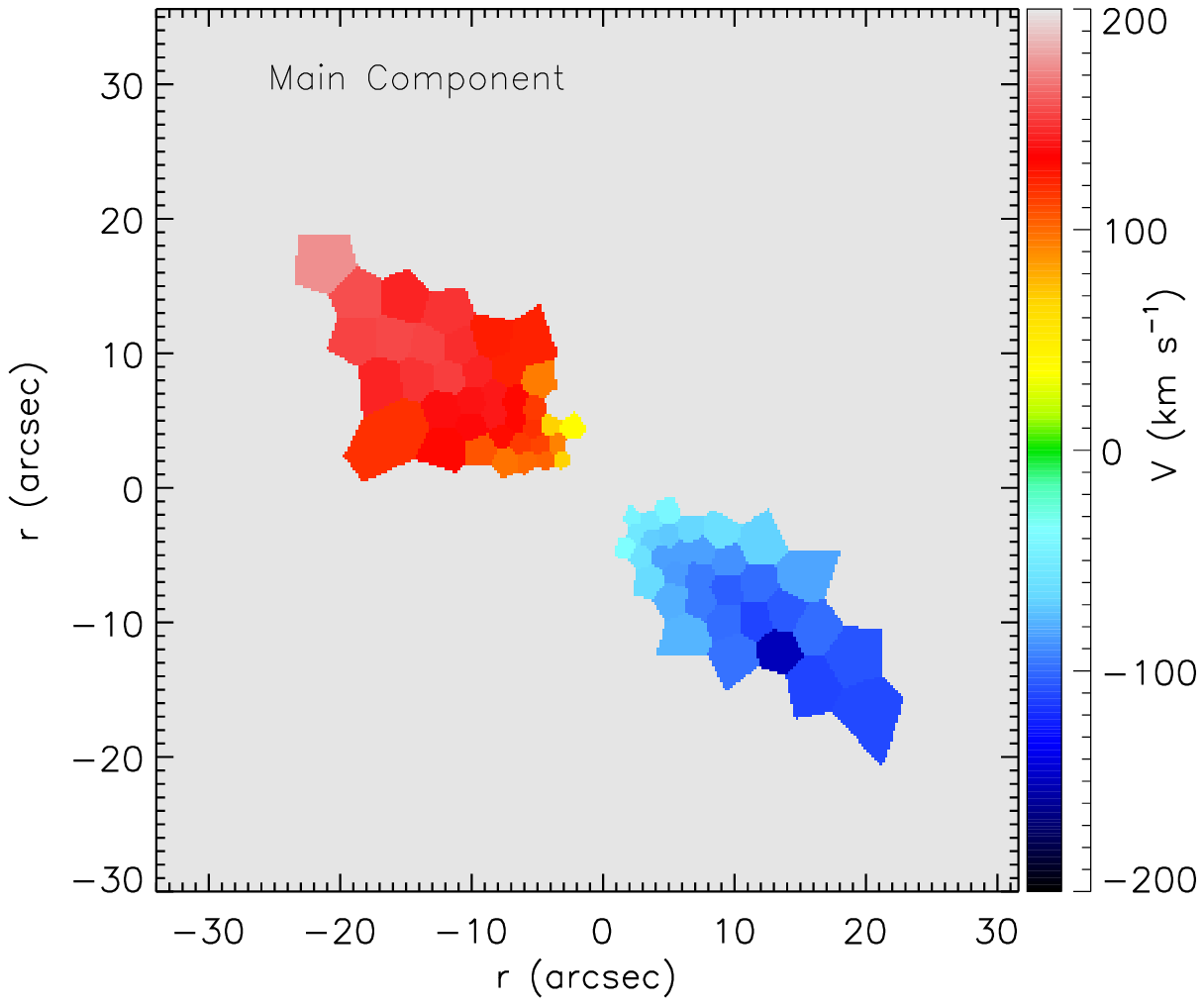}
\includegraphics[width=6.0cm]{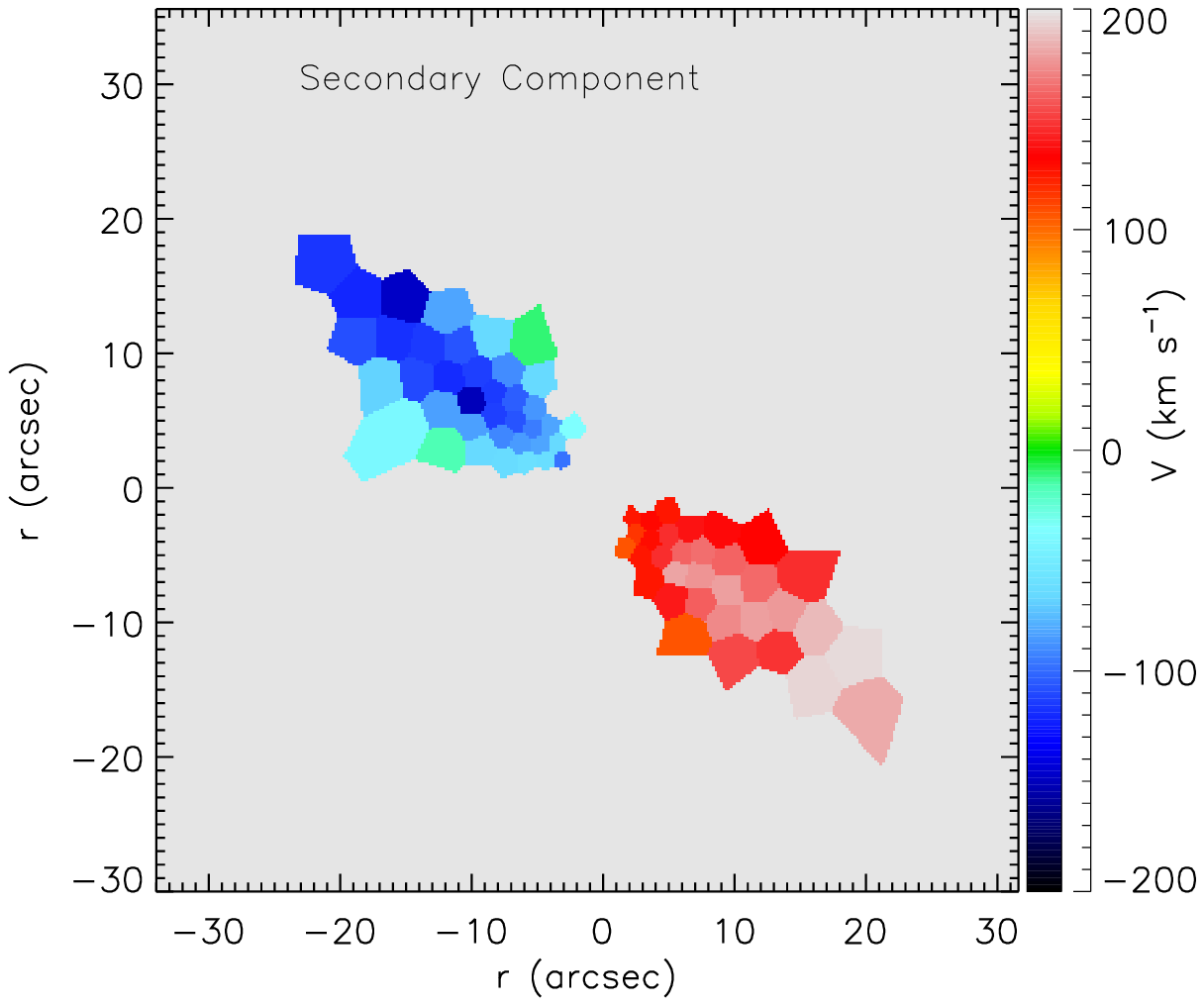}
\includegraphics[width=6.0cm]{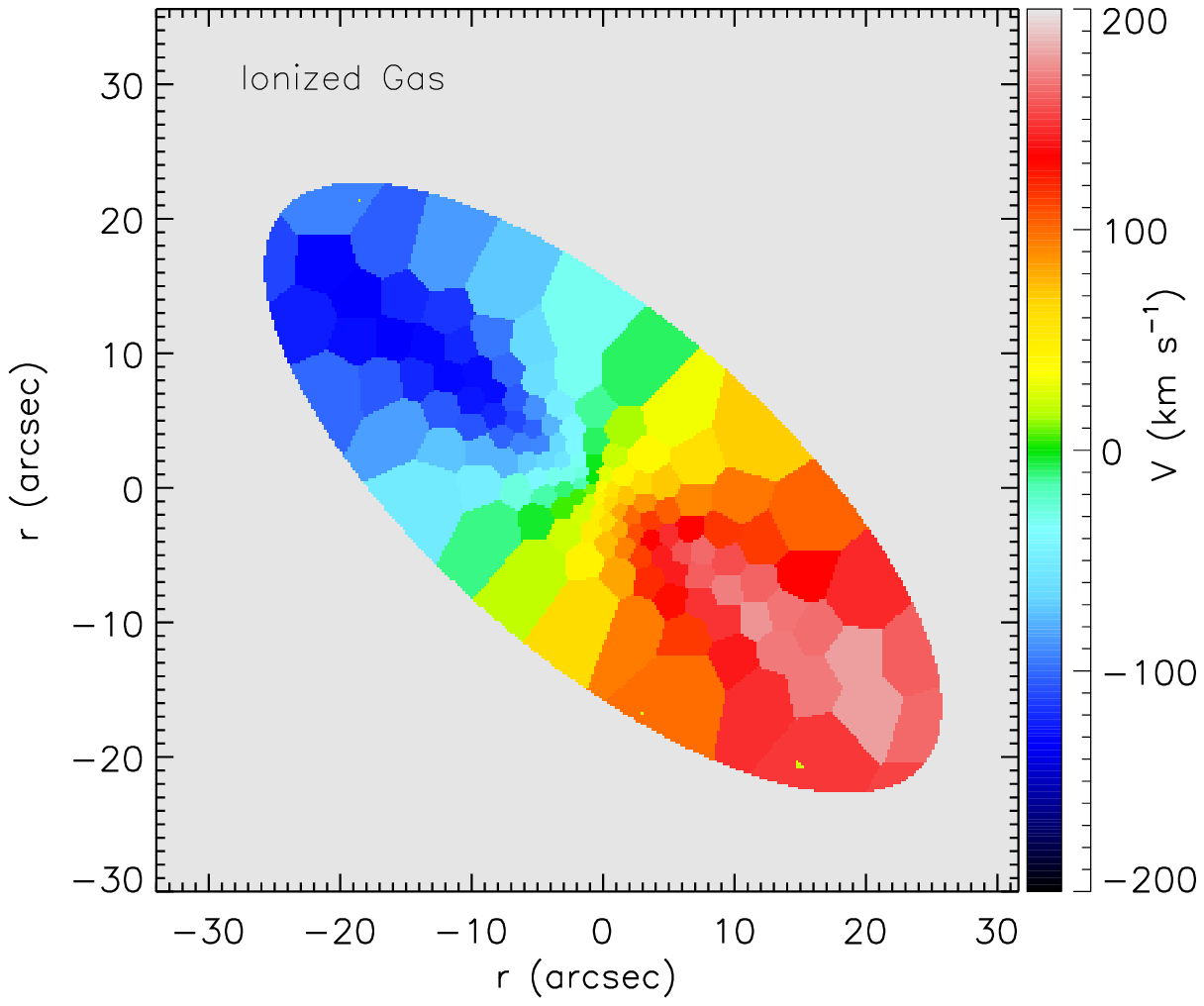}
\caption{Velocity field of the main component (left), secondary
  component (center), and of the ionized gas component
  (right). Typical errors are $7\,$\kms\ for the stellar components and
  $2\,$\kms\ for the ionized gas.}
\label{fig:vel}
\end{figure*}

\begin{figure*}
\centering
\includegraphics[width=6.0cm]{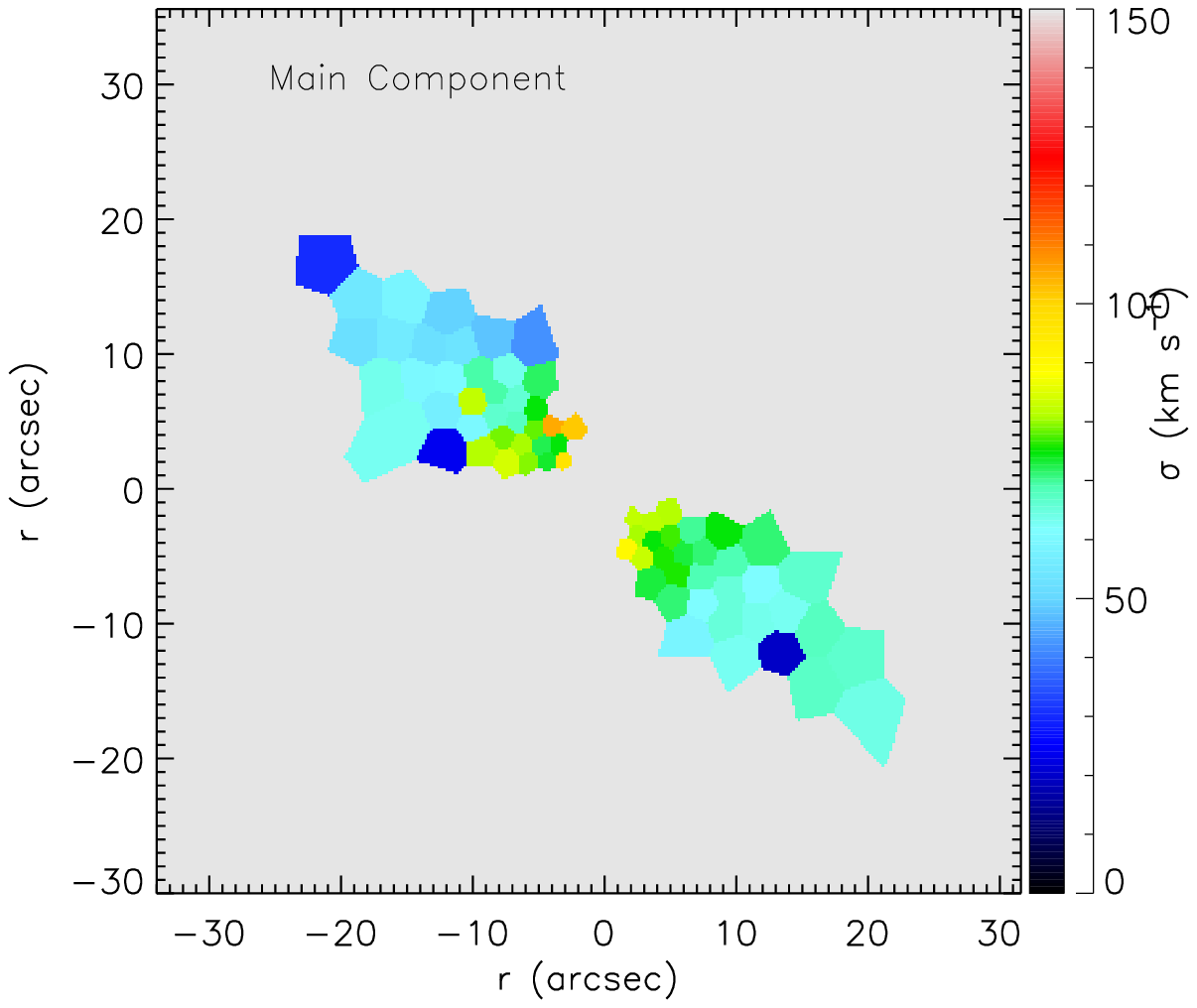}
\includegraphics[width=6.0cm]{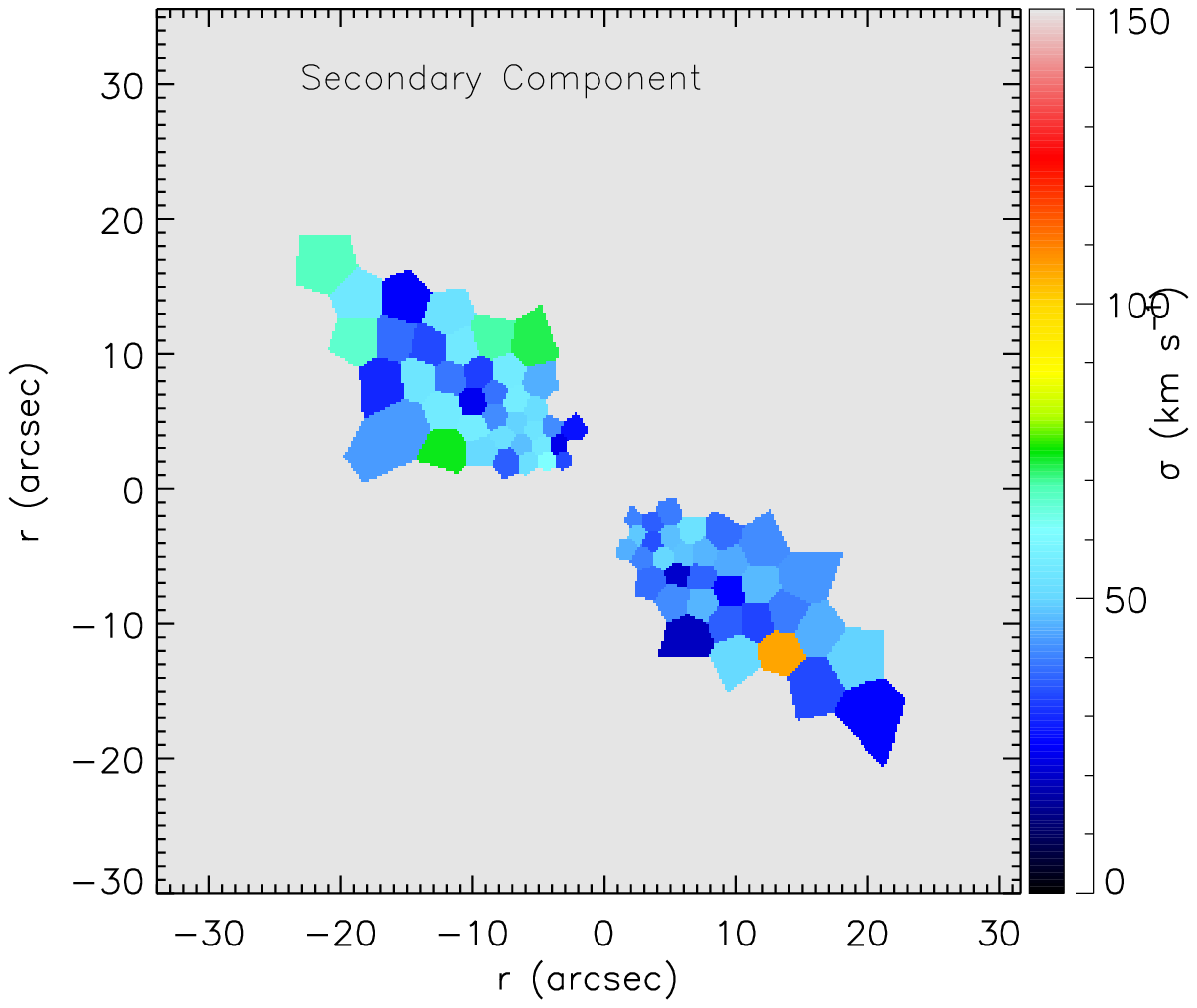}
\includegraphics[width=6.0cm]{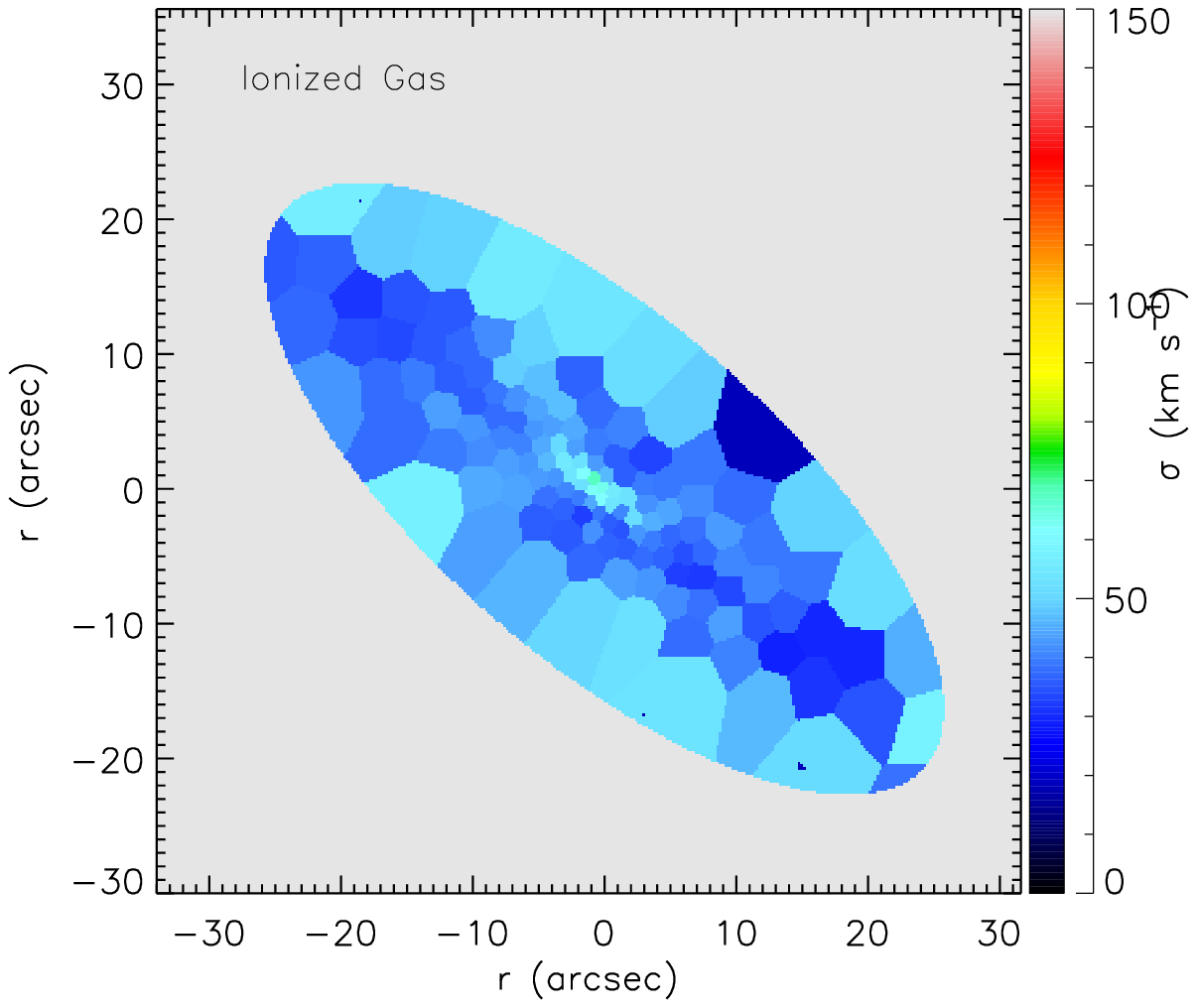}
\caption{As in Fig. \ref{fig:vel} for the velocity dispersion
  field. Typical errors are $10\,$\kms\ for the stellar components and
  $2\,$\kms\ for the ionized gas.}
\label{fig:disp}
\end{figure*}

After having measured all bins, we sort the two components using
their velocity.  We define the {\it main component} to be
the stellar component
which is receding on the NE side of the plane of the sky (and
therefore approaching on the SW side). This turned out to be the most
luminous component. We define the  {\it secondary component} to be
the other stellar component.
The decomposition is more robust in those
spatial bins where the velocity separation is high and the luminosity of
the two components is similar. We consider the two stellar components
to be ``kinematically separated'' if their velocity separation is
higher than $50\,$\kms, and $0.15 < Fr < 0.85$.
The spatial bins where we could not separate the two components identify
a circular region with radius $\sim$ 5\arcsec\ from the center
and a conic region aligned along the minor axis.
Consequently, the spatial bins
in which the stellar components can be considered as kinematically
separated identify a conic region aligned with the galaxy major axis.
From now on, although we fit the
kinematics on the whole galaxy, we consider only spatial bins
with the spectra of the two components well separated. Furthermore, we consider the whole galaxy only concerning the ionized gas emissions.
This choice minimizes issues related to degeneracies in the spectral
decomposition code (e.g., if two components have the same velocity,
the stellar templates used to determine the best fit can be assigned
to one component or the other without changing the best fitting
model), and the contamination of the galaxy bulge, which is confined
in the innermost $\sim$ 3\arcsec (see Fig. 1 in \citealt{2013ApJ...769..105K}).
In Figs. \ref{fig:vel} and \ref{fig:disp} we show the velocity and
velocity dispersion 2D fields of the two stellar and
ionized gas components, as measured in the CaT spectral region.

\begin{figure*}
\centering
\includegraphics[width=8.0cm]{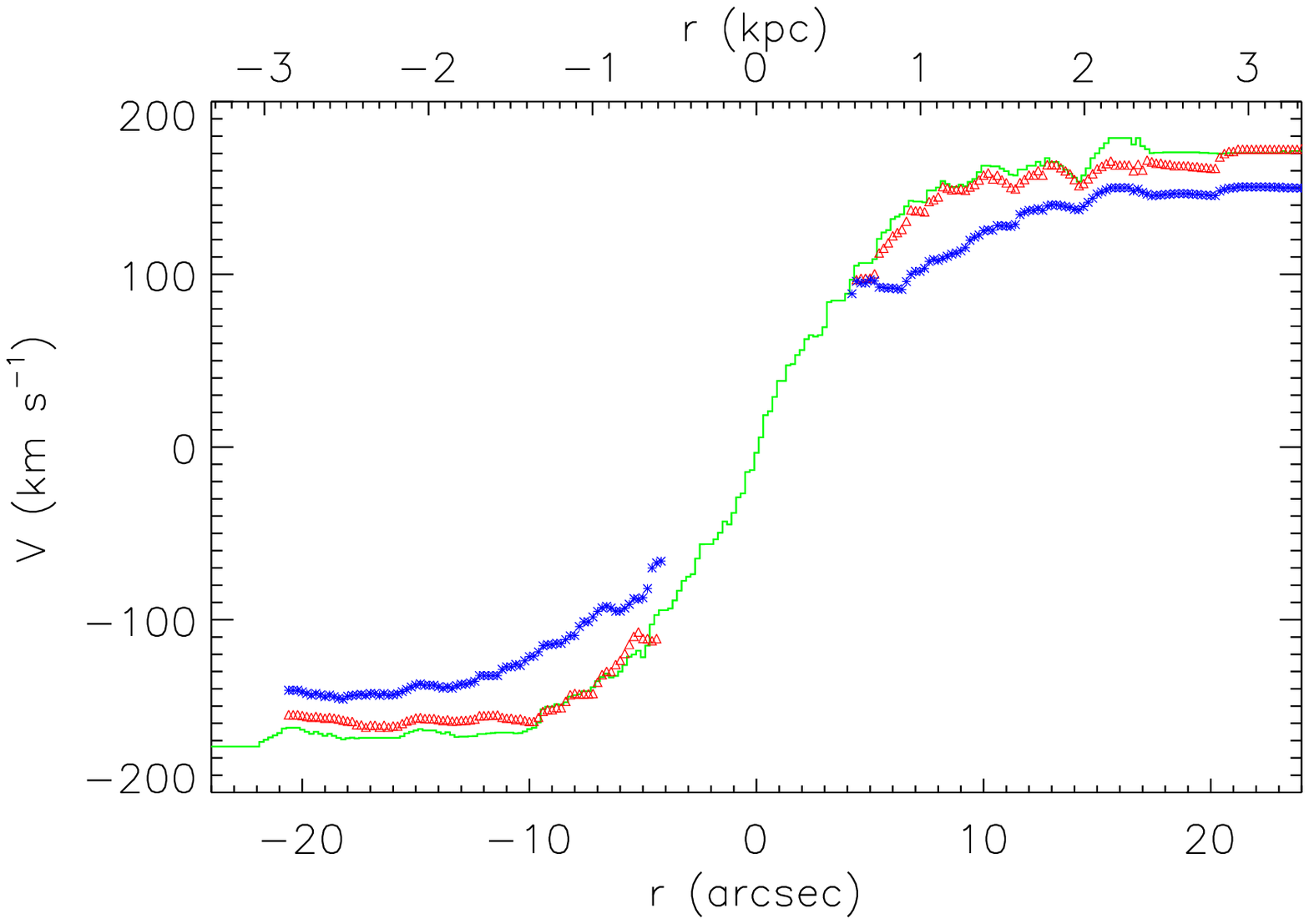}
\includegraphics[width=8.0cm]{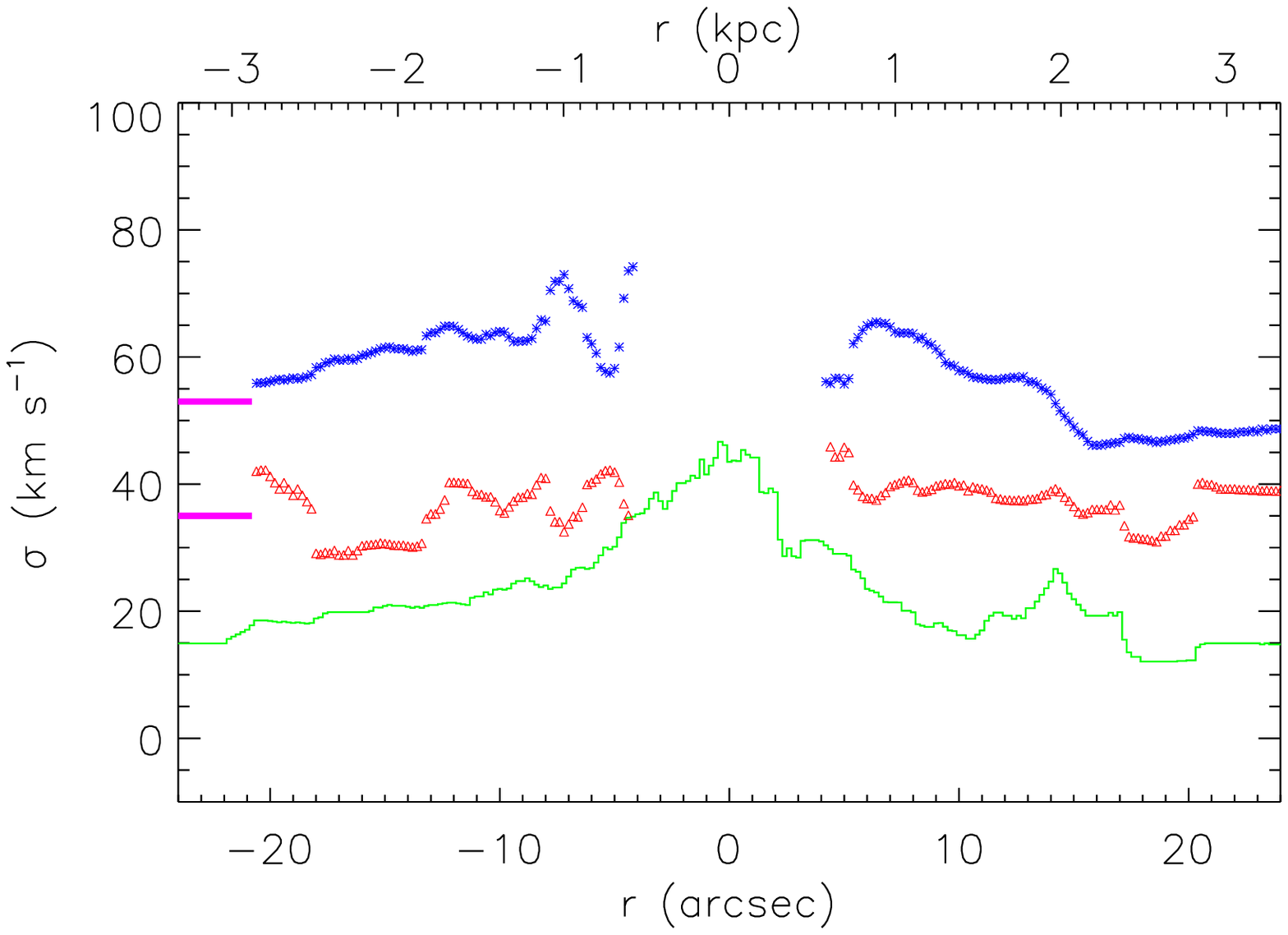}
\caption{Velocity (left) and velocity dispersion curve (right)
  extracted along the major axis of IC~719. Blue, red, and green lines
  represent the main stellar component, secondary stellar component,
  and ionized gas, respectively. The velocity of the main component
  has been changed in sign in order to allow a direct comparison with
  the velocity of the other two components. The instrumental velocity
  dispersion of $35\,$\kms\ (CaT) and $53\,$\kms\ (\Ha ) is indicated
  as reference by a horizontal tick. The extraction of the
  kinematics along the major axis has been done by spatial
    averaging the spaxels in Figs. \ref{fig:vel} and
   \ref{fig:disp} considering a $2$\arcsec -wide aperture passing
  thought the center along PA=$53^\circ$. Typical error-bars
    on the velocity and velocity dispersion are $5$ and $10\,$\kms , respectively.}
\label{fig:majaxis}
\end{figure*}
 
\subsection{Emission line maps}\label{sec:emlines}

\begin{figure*}
\centering
\includegraphics[width=8.0cm]{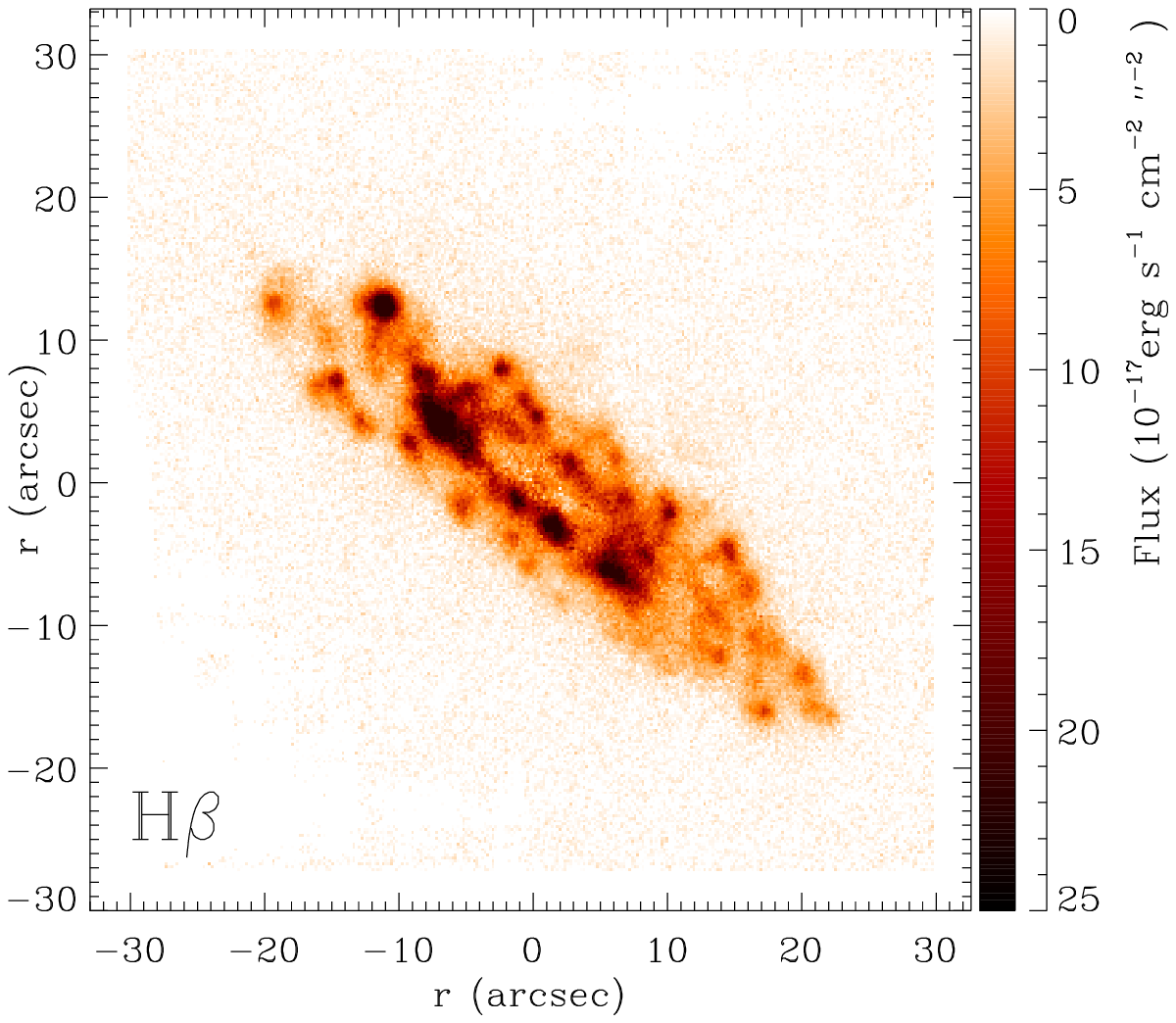}
\includegraphics[width=8.0cm]{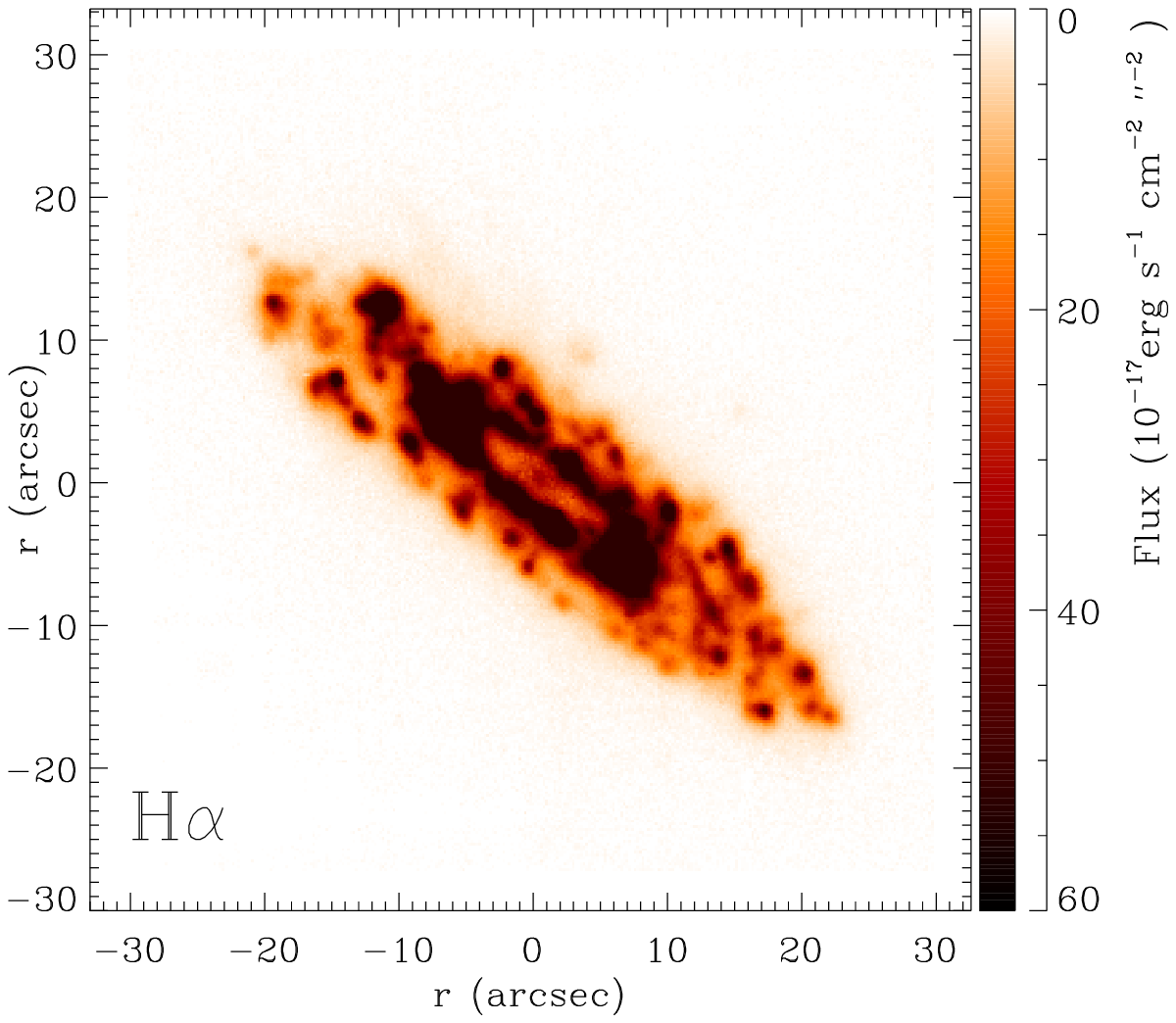}
\includegraphics[width=8.0cm]{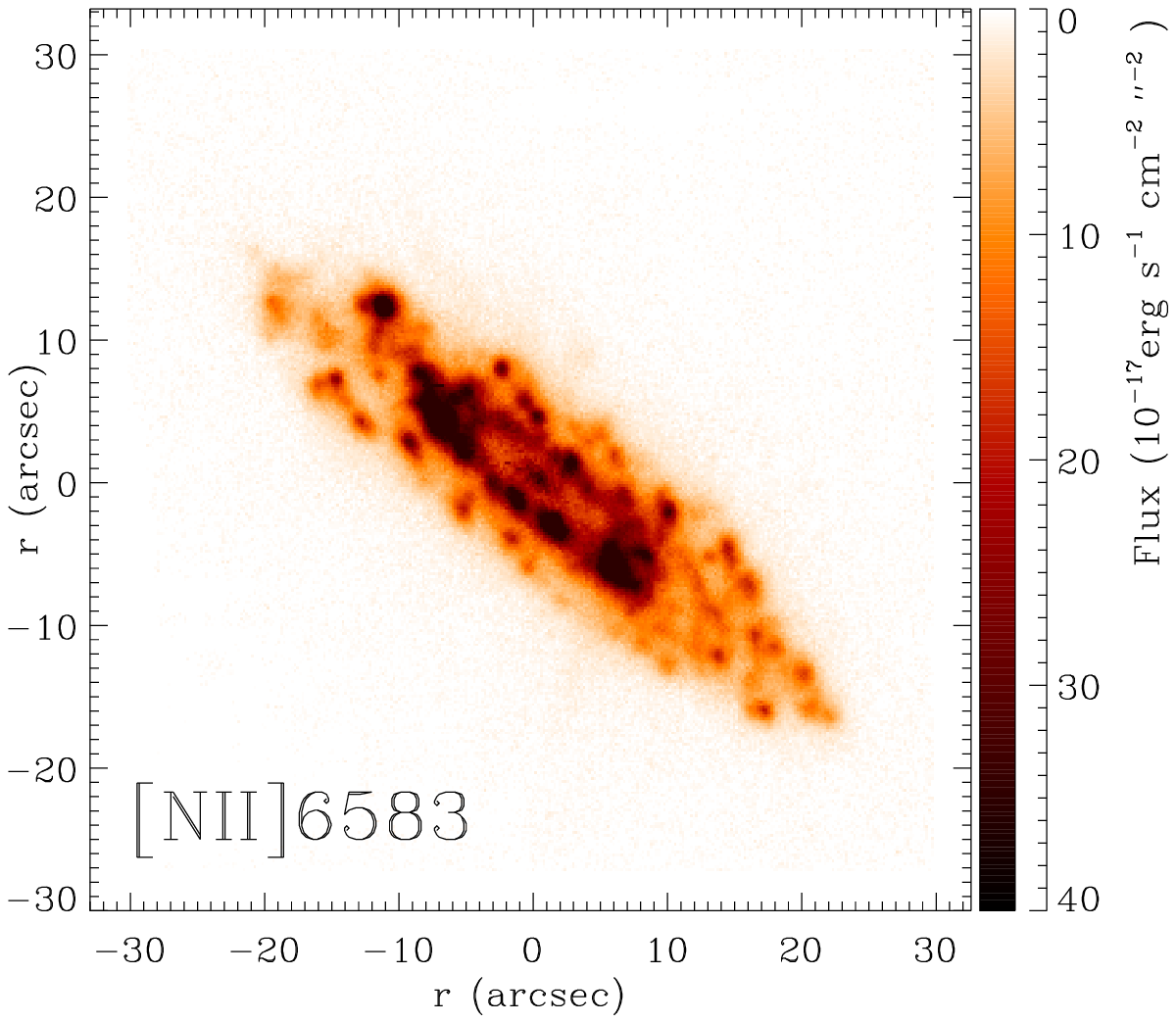}
\includegraphics[width=8.0cm]{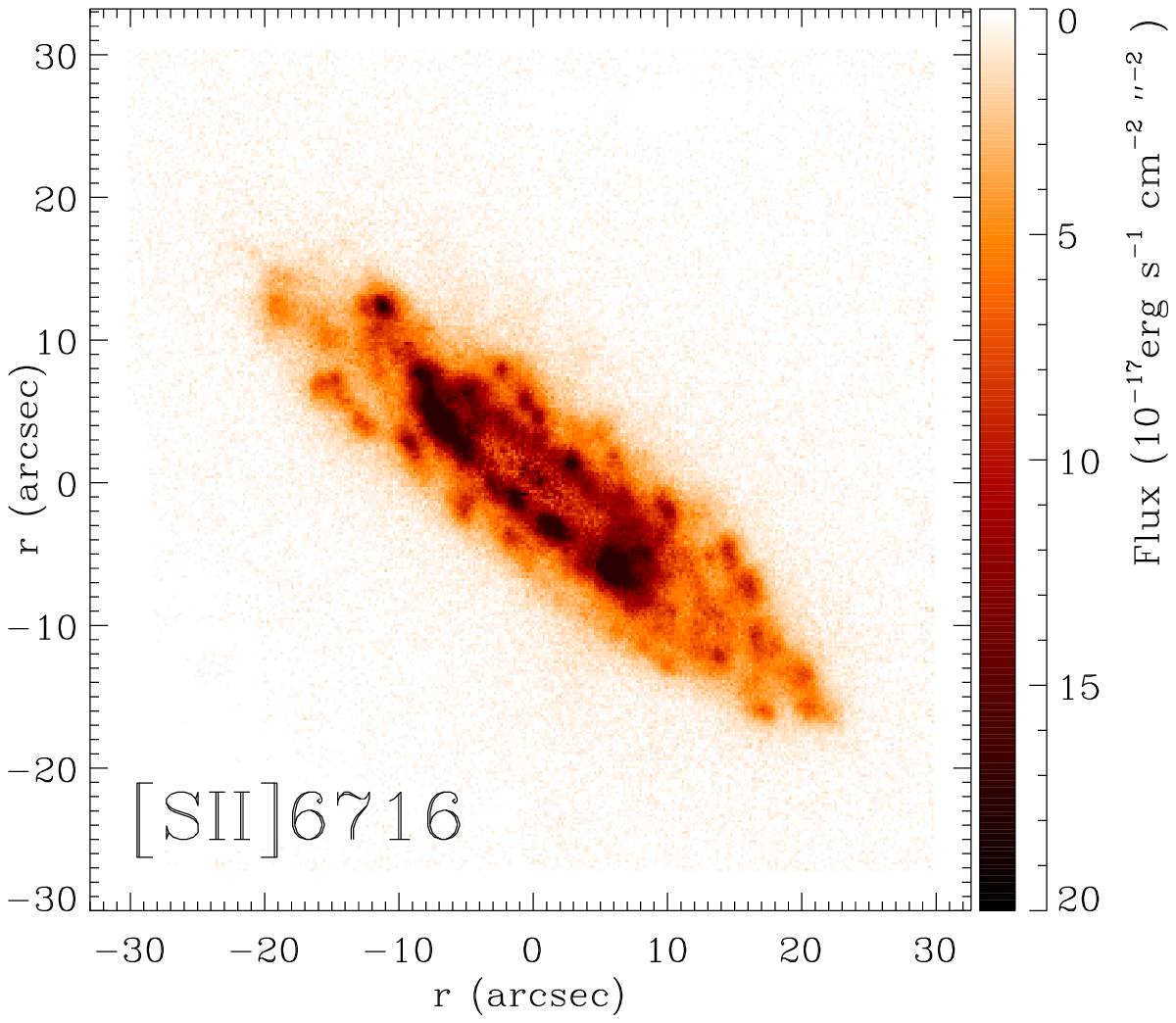}
\caption{Emission line maps, from top left to bottom right, for \Hb,
  \Ha, \niiig , and \siip . The field of view is
  $1'\times 1'$. N is up and E is left.  }
\label{fig:emlines}
\end{figure*}

After having derived the ionized gas velocity field, we extracted the
emission line maps of all the emission lines present in the spectral
range. The spectral lines, some of which were not included in the kinematical fit,
are \Hb , \oiiipg , \niipg , [He{\small
    I}]$,\lambda5875$, \oipg , \niiipg, \Ha, [He{\small
    I}]$,\lambda6678$, \siipg, and [Ar{\small I}]$,\lambda7135$.  We
do this on the unbinned datacube in order to draw maps with the
original spatial sampling of the instrument.  At each spaxel of the
unbinned datacube, we subtract the stellar best-fitting spectrum of
the Voronoi bin belonging to that specific spaxel. The stellar model
spectrum was scaled by a factor that was chosen such that it matches
the stellar continuum. The wavelength of the emission line is computed
from the ionized gas velocity measured in the same Voronoi bin. Subsequently,
for each spaxel, the emission line flux is obtained by integrating the
signal of an
$11\,$\AA-wide  wavelength region centered on the line. The width of $11\,$\AA\ was used in order to include
all the flux of the line and to avoid contamination from the nearby
emission lines. This produces an emission line map with a spatial
sampling of $0\farcs 2$.
Examples of the four main emission-line
maps are shown in Fig. \ref{fig:emlines}.

\begin{figure*}
\centering
\includegraphics[width=4.0cm]{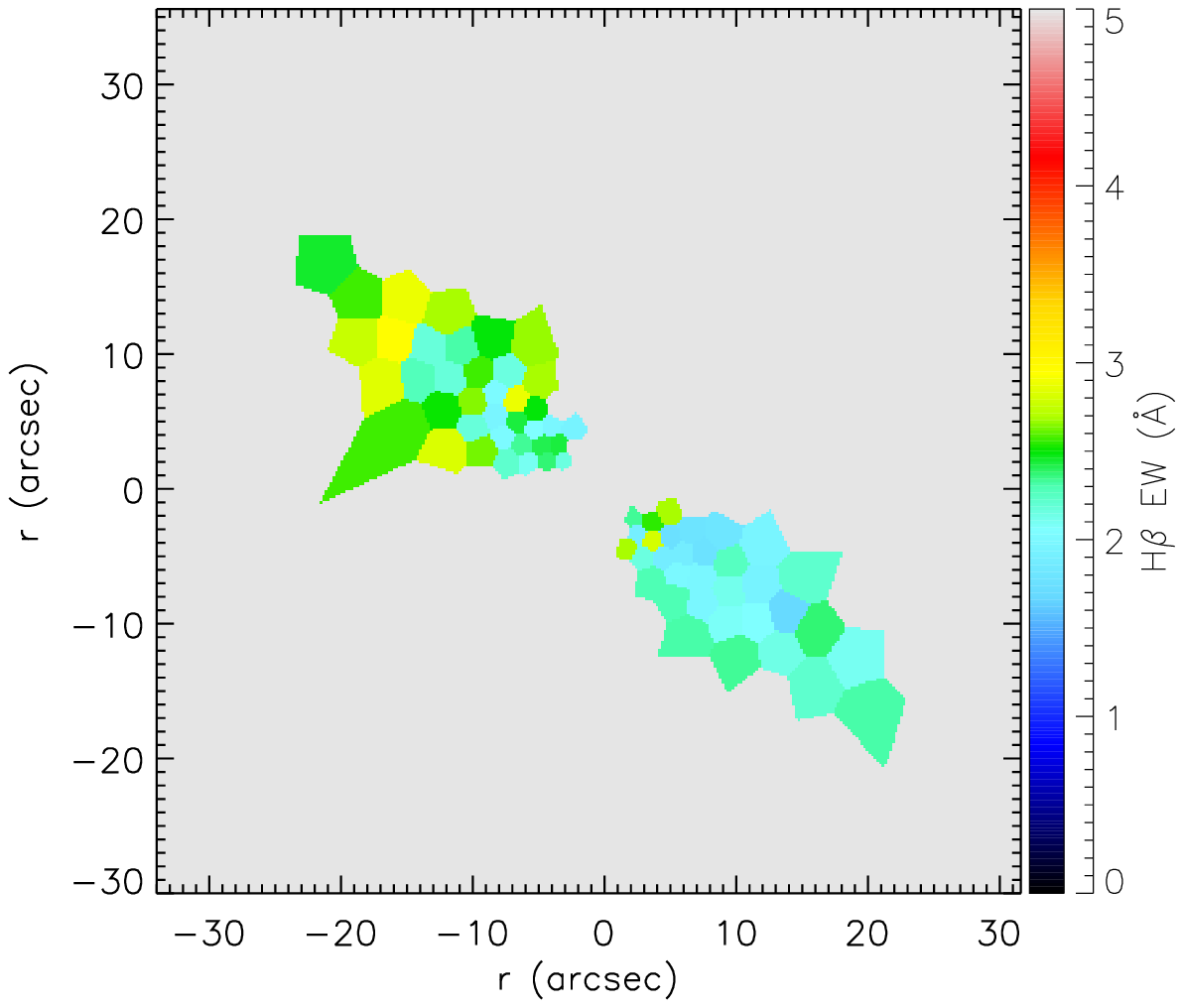}
\includegraphics[width=4.0cm]{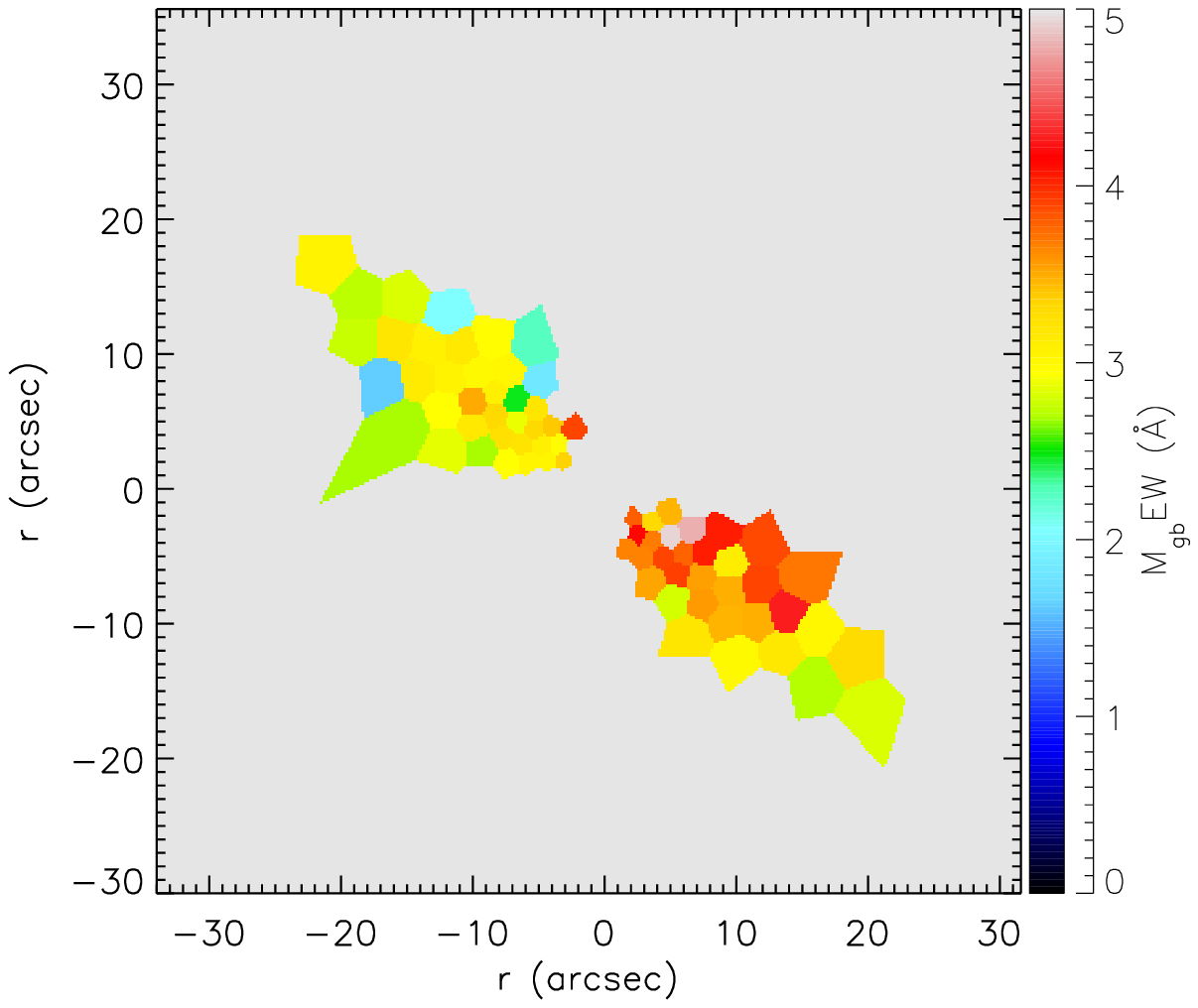}
\includegraphics[width=4.0cm]{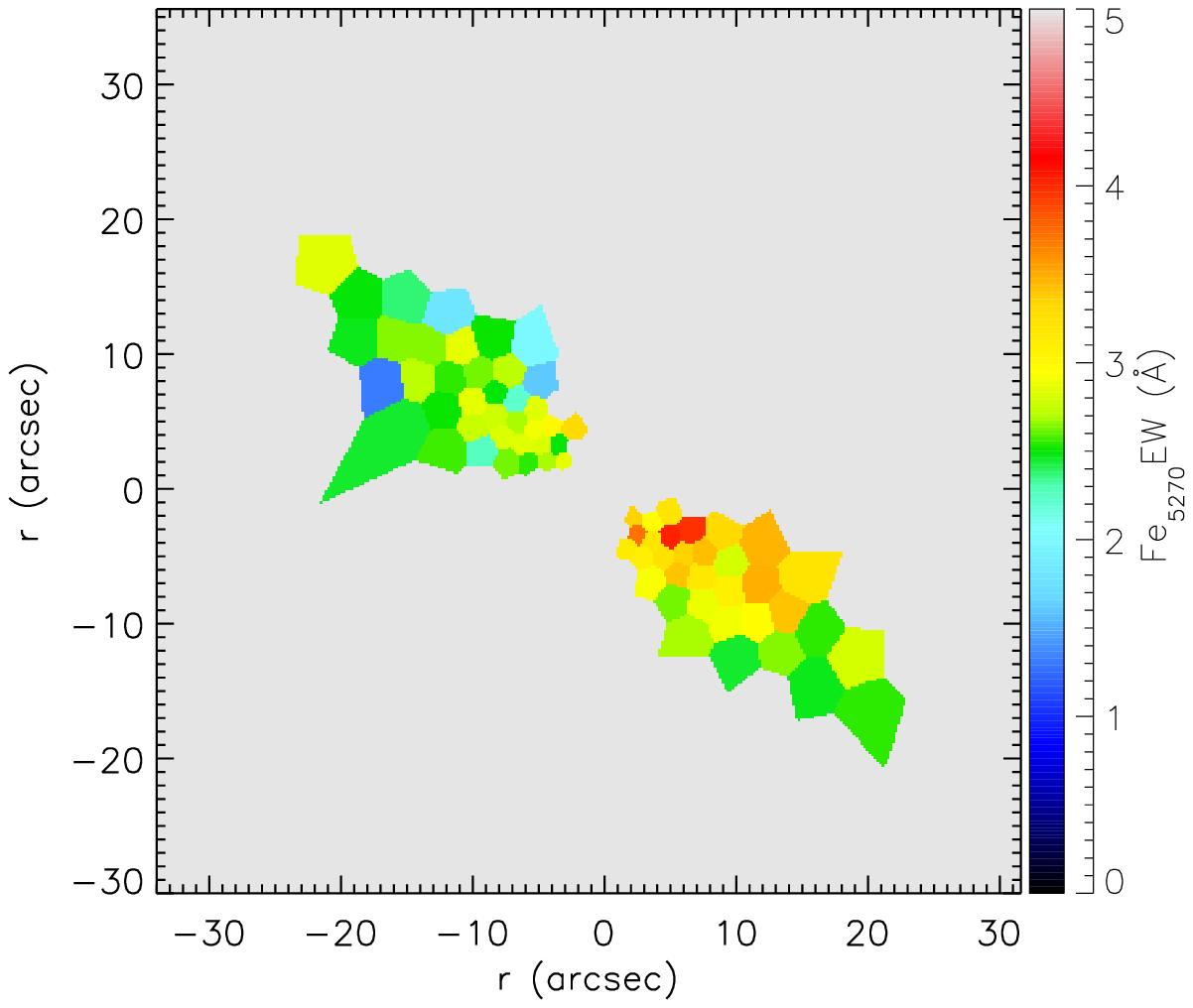}
\includegraphics[width=4.0cm]{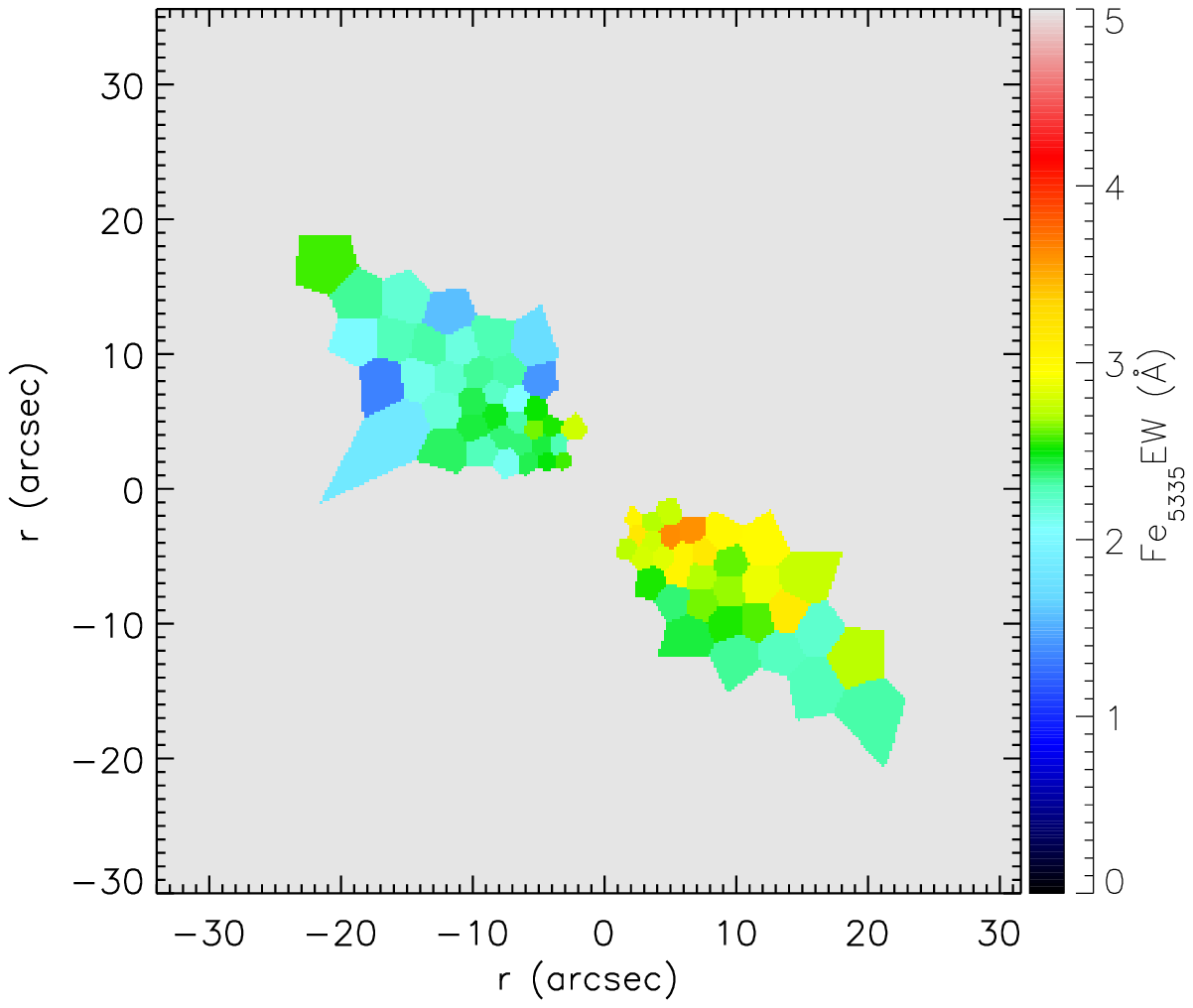}
\caption{Equivalent width map of the main component for the \Hb, \Mgb,
  {\rm Fe}$_{5270}$, and {\rm Fe}$_{5335}$ absorption lines. The color
  indicates the EW in each bin as indicated by the color bar.  The
  field of view is $1'\times 1'$. N is up and E is left.  }
\label{fig:EWmaps1}
\centering
\includegraphics[width=4.0cm]{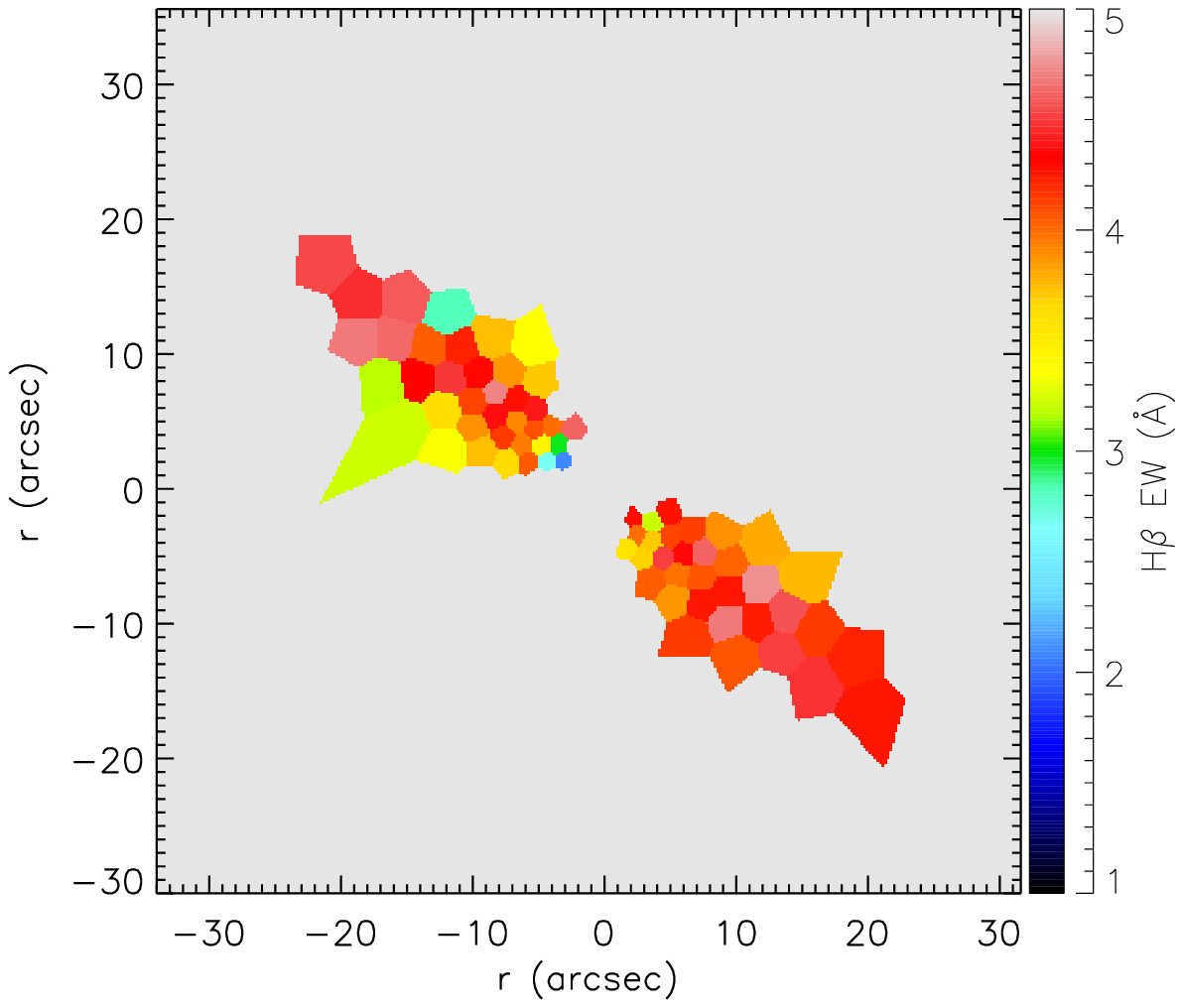}
\includegraphics[width=4.0cm]{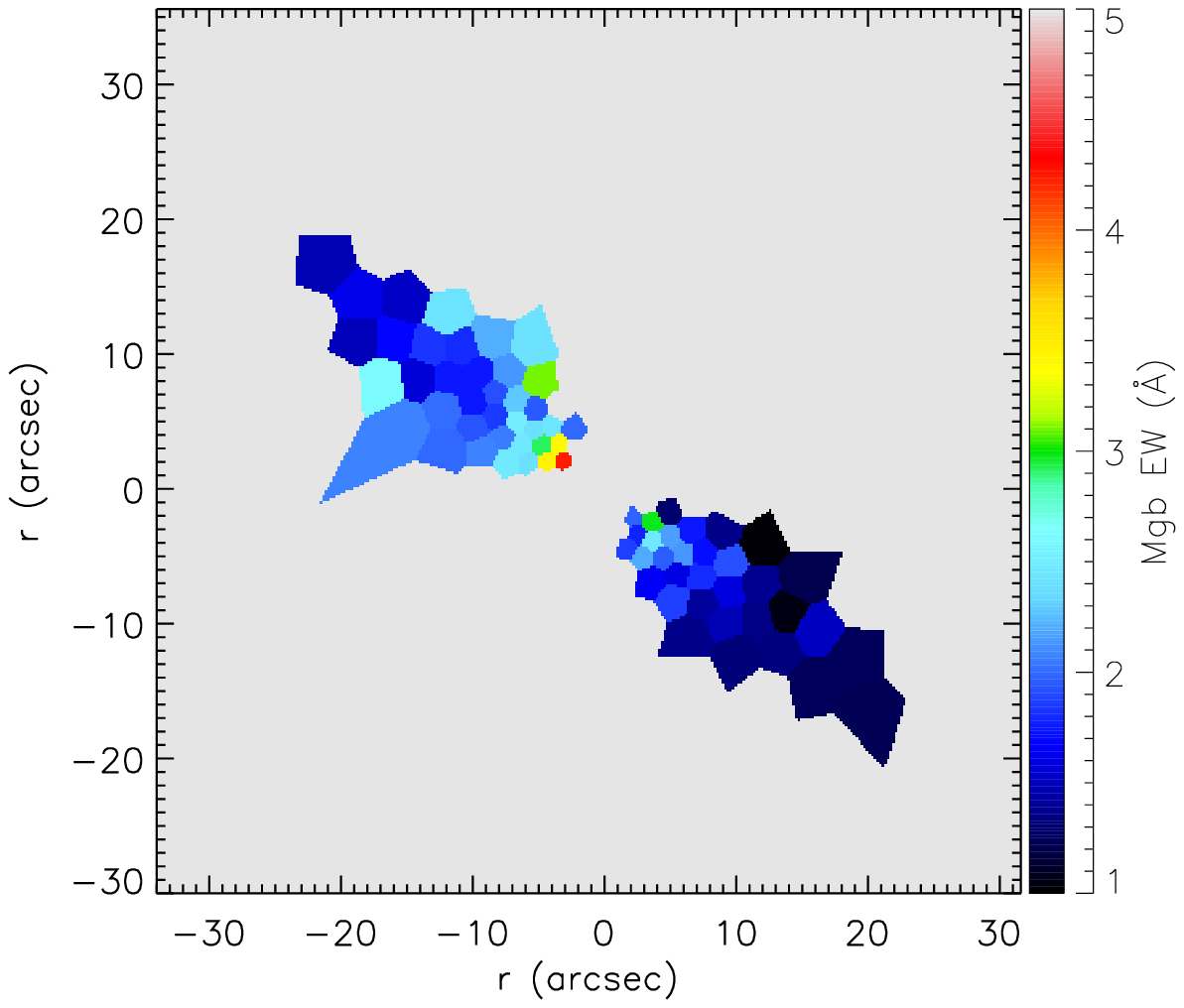}
\includegraphics[width=4.0cm]{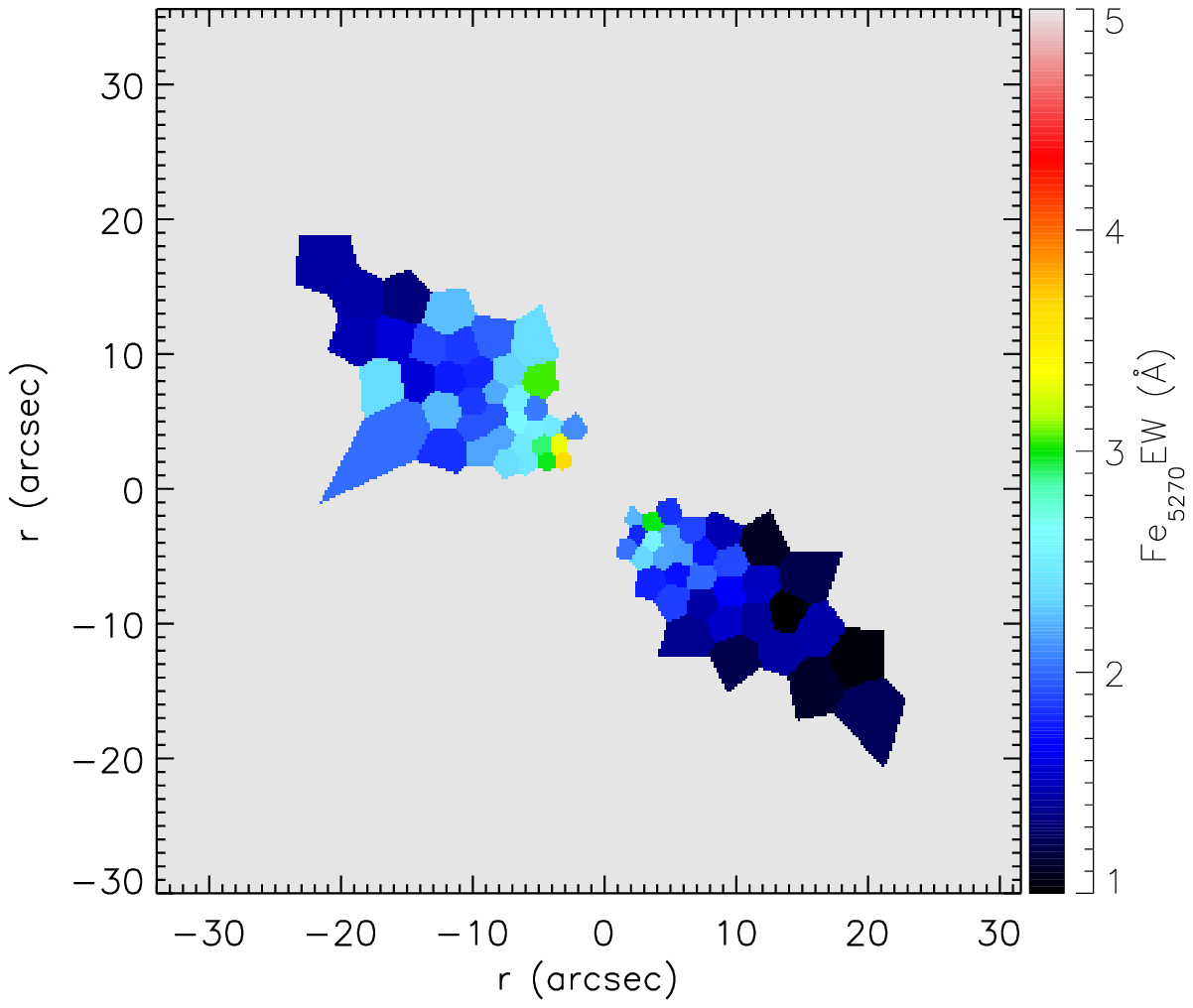}
\includegraphics[width=4.0cm]{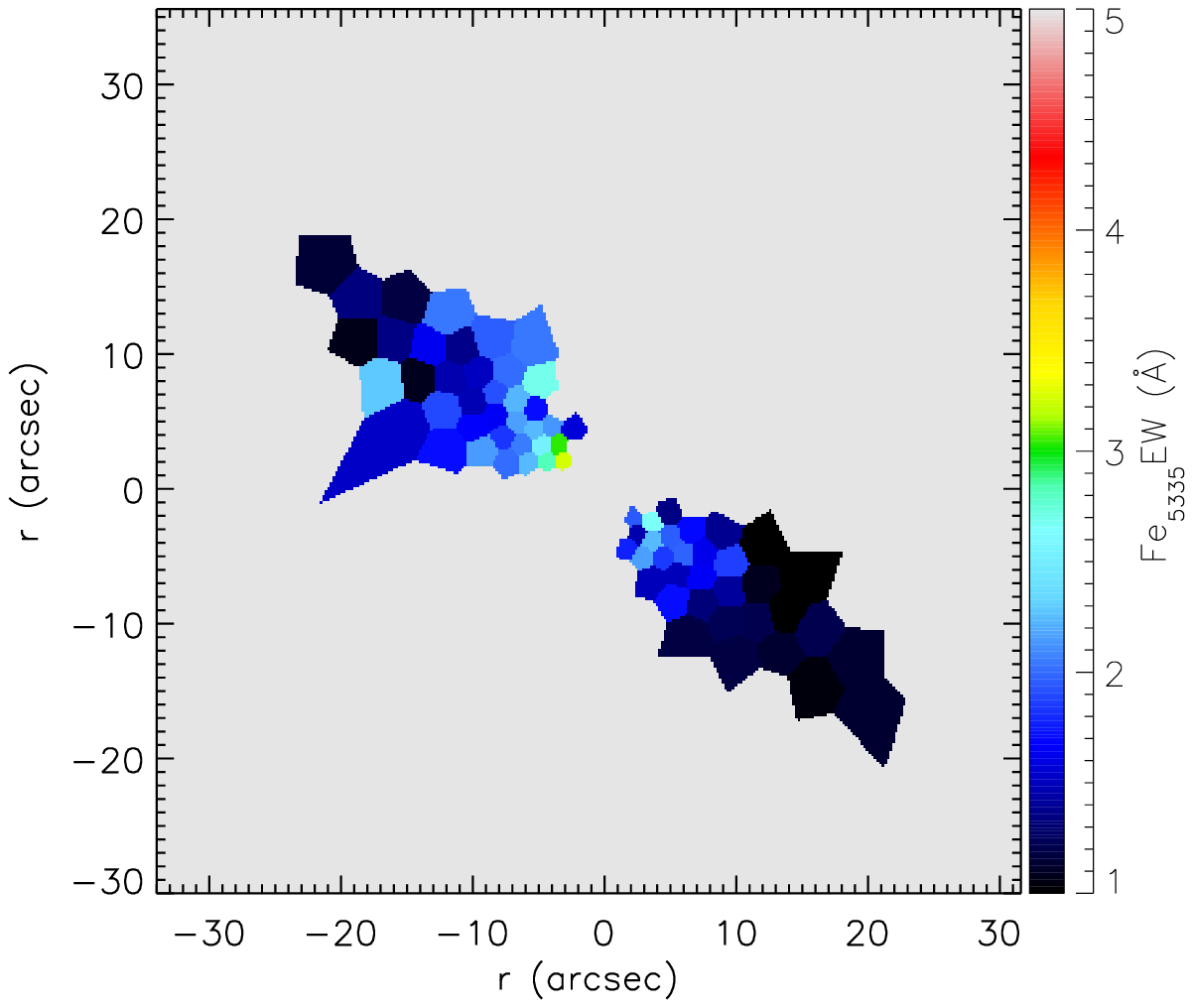}
\caption{As in Fig. \ref{fig:EWmaps1} but for the secondary component.}
\label{fig:EWmaps2}
\end{figure*}

\subsection{Line-strength indices}\label{sec:lick}
\begin{figure*}
\centering
\includegraphics[width=8.0cm]{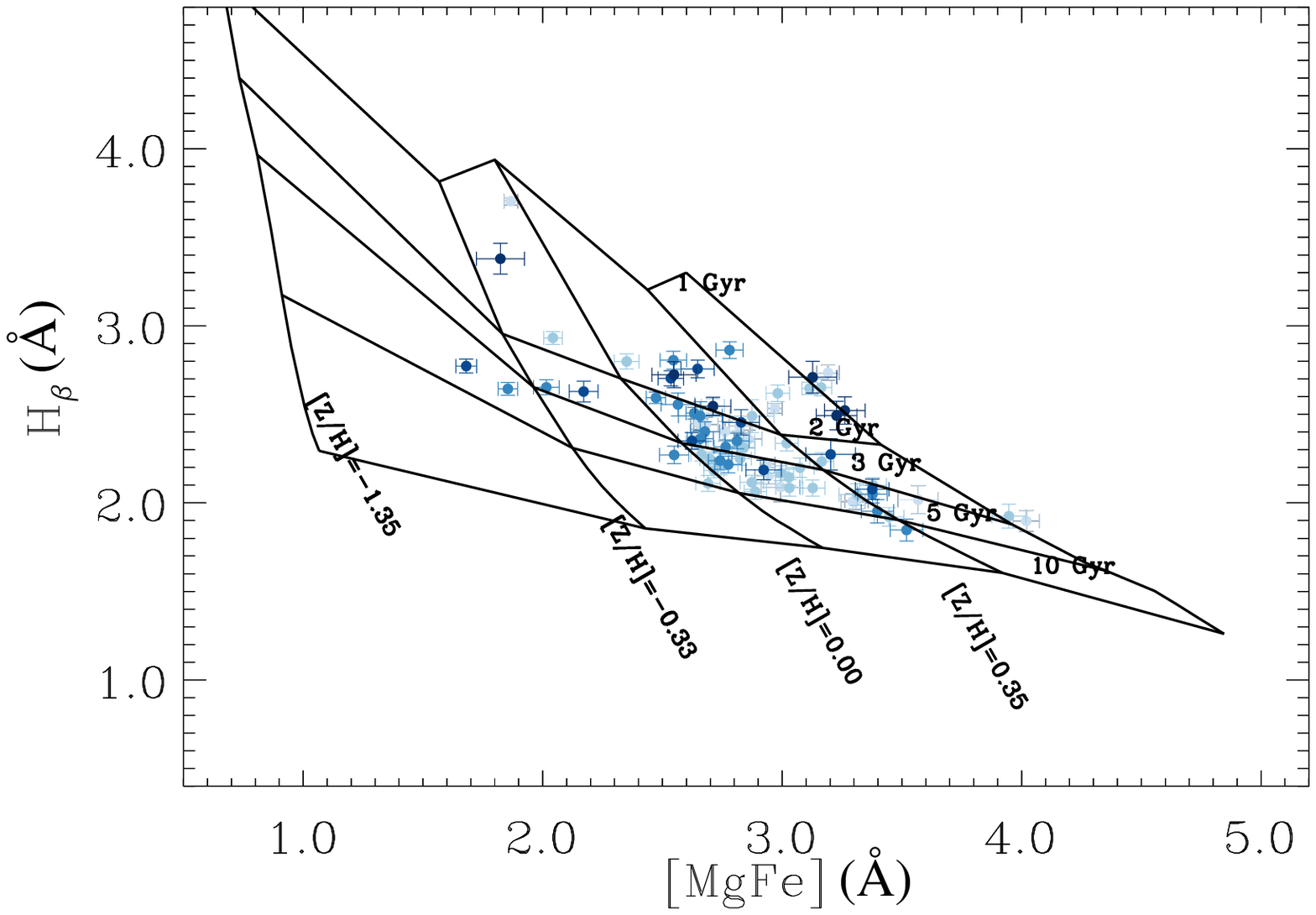}
\includegraphics[width=8.0cm]{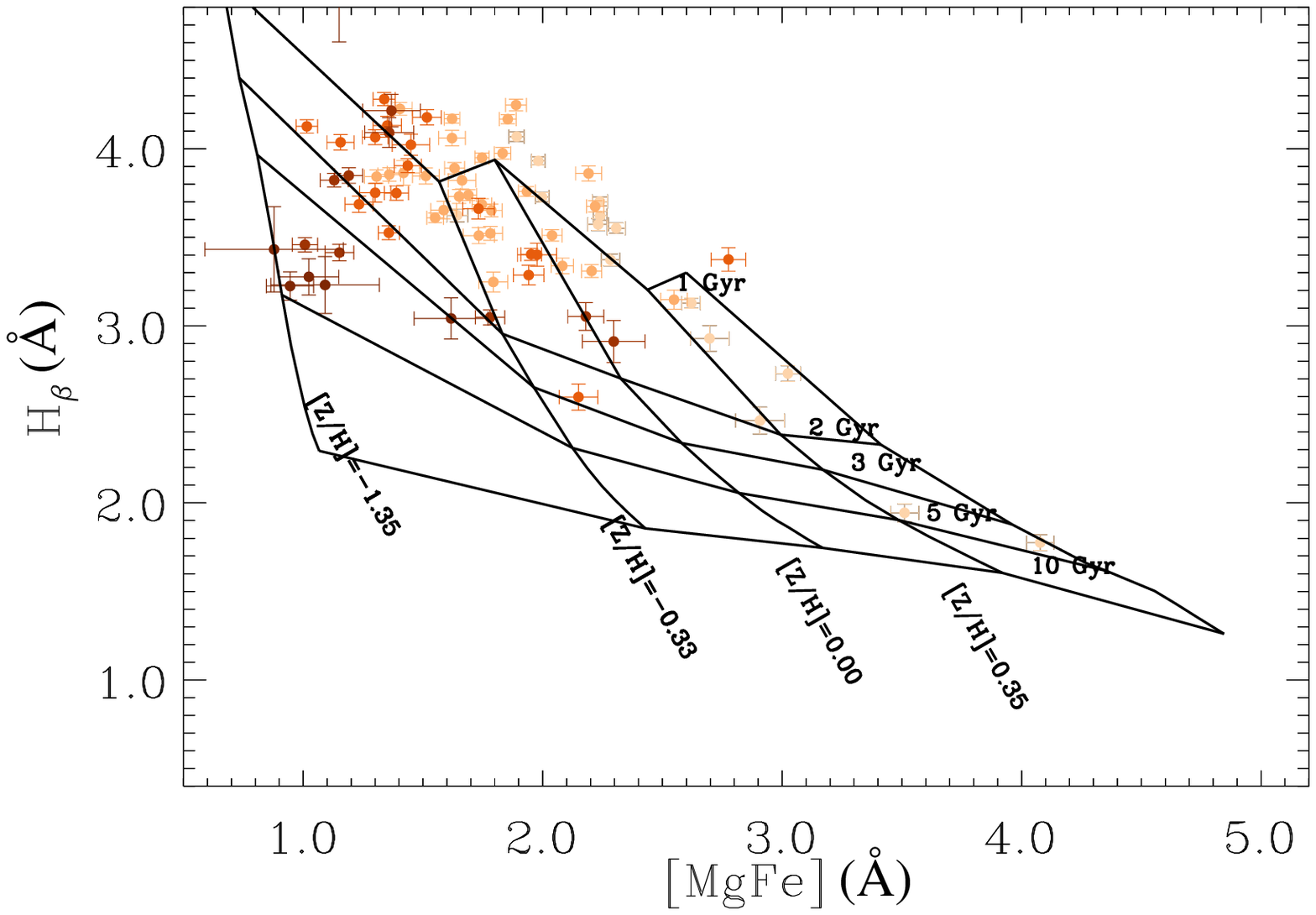}
\caption{Lick indices for the main component (left panel, blue dots) and the
  secondary component (right panel, red dots). The color tonality indicates the
  position of the measurement with lighter colors for the central
  regions and darker colors for the external regions.  The lines are
  the loci of constant age and metallicity as indicates by the
  labels.}
\label{fig:lick1}
\end{figure*}

We measured the Lick indices on the best fit spectra obtained in the
spectral fit (Fig. \ref{fig:specfit_hb}, blue and red lines).  We
show in Figs. \ref{fig:EWmaps1} and \ref{fig:EWmaps2} the 2D maps of the equivalent
width (EW) of the \Hb, \Mgb, {\rm Fe}$_{5270}$, and {\rm Fe}$_{5335}$
absorption lines.
The combined magnesium-iron index
$[{\rm MgFe}]^{\prime}=\sqrt{{\rm Mg}\,b\,\left(0.72\cdot {\rm Fe5270} +
  0.28\cdot{\rm Fe5335}\right)}$ \citep{2003A&A...401..429T} and the
average iron index
$\langle$Fe$\rangle = \left( {\rm Fe}_{5270} +{\rm Fe}_{5335}\right)/2$ were
computed too. Errors on indices were derived from photon statistics
and CCD readout noise, and calibrated by means of Monte Carlo
simulations \citep{2012MNRAS.423..962M, 2016MNRAS.463.4396M}.
In Fig. \ref{fig:lick1}, we compare the measured line-strength indices to the
prediction of simple stellar population models by \citet{Thomas+11}.
The two stellar components occupy different areas of the
diagnostic plots of Fig.  \ref{fig:lick1} and show a wide range of
metallicities and ages. 

%

\section{Results}
\label{sec:results}

 The spectral decomposition successfully identified the presence of
  two stellar components in IC~719 (Figs. \ref{fig:specfit_CaT} and
  \ref{fig:specfit_all}).
We determined the rotation curves and the velocity dispersion radial
profiles (Sect. \ref{sec:res:kinem}) and the stellar populations of the
two components (Sect. \ref{sec:ssp}).
In Sect. \ref{sec:distribution}, we reconstruct the images of the two
stellar disks and study their morphology; in Sect. \ref{sec:sfr},
we use the measured emission lines to derive the spatially resolved
SFR.

\subsection{Kinematics}
\label{sec:res:kinem}

The spectral decomposition confirmed the presence of two extended
stellar components in IC~719 that rotate in opposite directions, as
shown in Fig. \ref{fig:vel}. The larger the velocity difference, the
clearer the separation between the absorption lines of the two
stellar components. The double-peaked feature of the absorption lines
is particularly evident in the CaT region, where the
instrumental velocity resolution is better than in the Mg region
(Figs. \ref{fig:specfit_CaT} and \ref{fig:specfit_hb}). In comparison with
previous works based on the Mg region, the use of the CaT not only
allows a more precise determination of the velocity, but also a more
reliable measurement of the velocity dispersion, which is important to
infer the formation scenario (see Sect. \ref{sec:Discussion}).

The main stellar component, which is the one receding in the NE side
of the galaxy, reaches a rotation of $\pm 150\,$\kms\ at about
$2\,$kpc from the galaxy center and has an almost constant rotation
velocity out to the outermost observed bin ($\approx 3\,$kpc). Its
velocity dispersion ranges from $\sim 70\,$\kms\ in the most central
radius we could measure ($0.5\,$kpc) to $\sim 50\,$\kms\ in the
outermost regions.  The secondary stellar component rotates in the
opposite direction, and is characterized by a slightly steeper
velocity gradient and higher rotation amplitude: $180\,$\kms\ reached
at $\approx 2\,$kpc from the galaxy center. The secondary stellar
component is kinematically colder than the main component, with a
velocity dispersion that ranges between $50\,$\kms\ at $\sim 0.5\,$kpc
and $30\,$\kms\ in the outer regions, a value smaller than the
instrument velocity resolution.

The ionized gas rotates along the same direction as the
secondary stellar component; its rotation amplitude and radial profile
of the velocity dispersion are similar to those of the
secondary stellar component (Fig. \ref{fig:majaxis}).

\subsection{Morphology of the stellar disks}
\label{sec:distribution}

\begin{figure*}
\centering
\includegraphics[width=6.0cm]{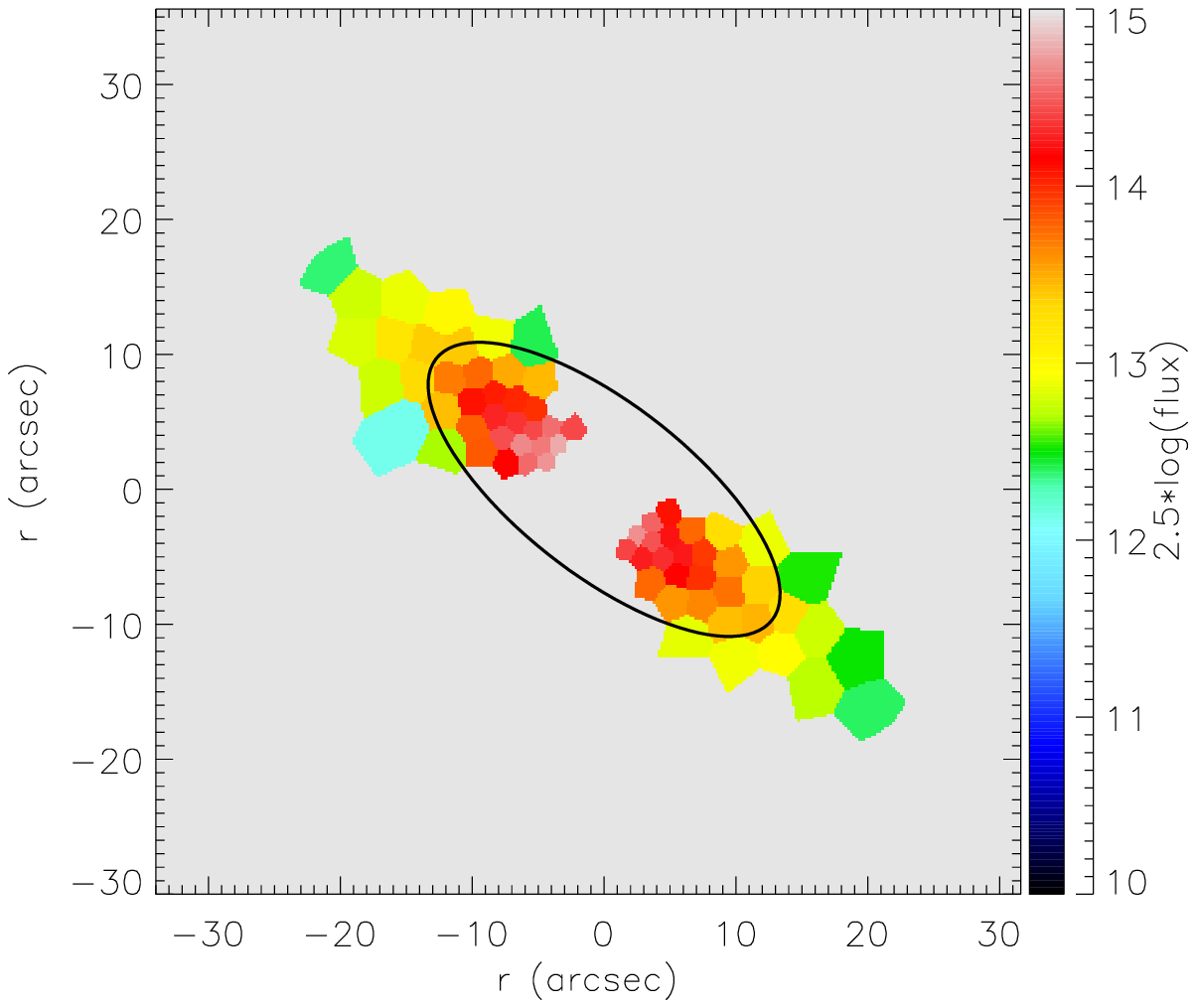}
\includegraphics[width=6.0cm]{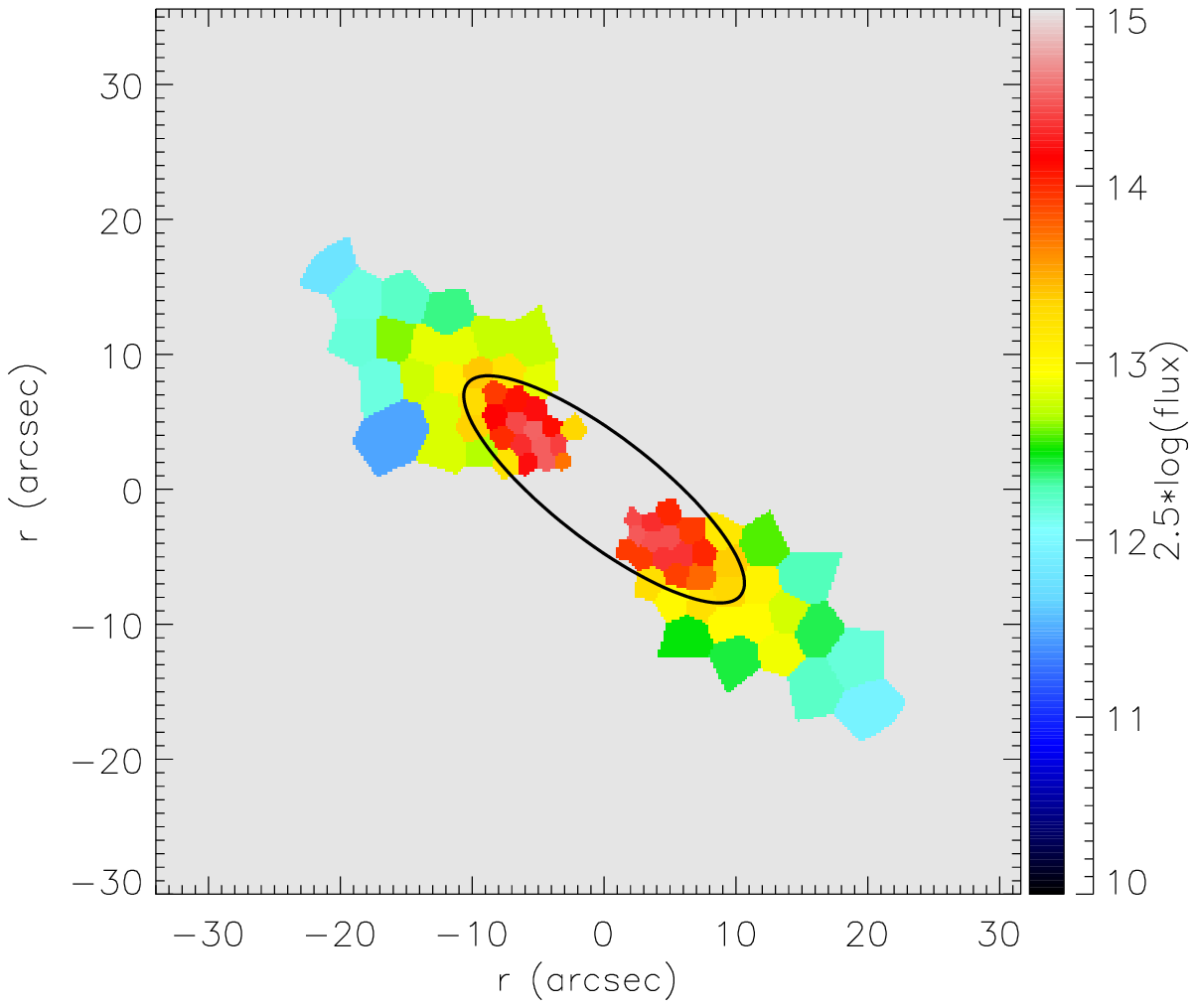}
\caption{Image of the primary (left) and secondary components (right)
  reconstructed from the relative flux contributions. The black line
  represents an ellipse with ellipticity of 0.6 and 0.7 for the
  primary and secondary components, respectively.}
\label{fig:ima_2comp}
\end{figure*}

We derive the 2D light distributions of the two stellar
disks, their surface brightness profiles, and their relative flux
contributions from the $Fr$ measured at each spatial bin by the
spectral decomposition code and the galaxy image. We applied
  the same Voronoi binning used for the kinematical analysis to the
  reconstructed image of IC 719. We then obtained the image of the
  primary stellar component by multiplying it by the $Fr$ map (or the
  secondary component by multiplying it by $1-Fr$).
Figure \ref{fig:ima_2comp} shows the reconstructed images of the two
stellar components. We can then derive the radial surface brightness
profile of both stellar components (Fig. \ref{fig:radprof}). This
  is done using the task {\it ellipse} of the {\it IRAF}\footnote{IRAF
    is the Image Reduction and Analysis Facility, written and
    supported by the National Optical Astronomy Observatories (NOAO).}.
The main component has an exponential
profile with scale length of $\sim1.5\,$kpc; the secondary component
follows an exponential profile with a possible truncation beyond
$4\,$kpc. Its scale length is $\sim1\,$kpc and therefore smaller than
the primary disk. We also used the reconstructed images to
derive the isophotal shape at different radii of the two stellar
components. Because of the limited number of data points, we first
created a regular square matrix of pixels with the same size as the
MUSE field of view. We assigned to each pixel the corresponding flux
value belonging to that Voronoi bin. We then fitted ellipses with
fixed center and position angle to this frame. The result is plotted
in Fig. \ref{fig:radprof}.

Both the main and secondary components contribute 
$(50\pm15)\%$ of the stellar luminosity at about $1.0\,$kpc, while the
main component dominates in the outer regions of the galaxy being
$(65\pm15)\%$ at about $\simeq 2.5\,$kpc, and $(85\pm15)\%$ at about
$\simeq 6.0\,$kpc. In total, the secondary component contributes to $35$\%
of the galaxy light as measured from $0.8$ to $4.2\,$kpc.

\begin{figure}
\centering
\includegraphics[width=8.0cm]{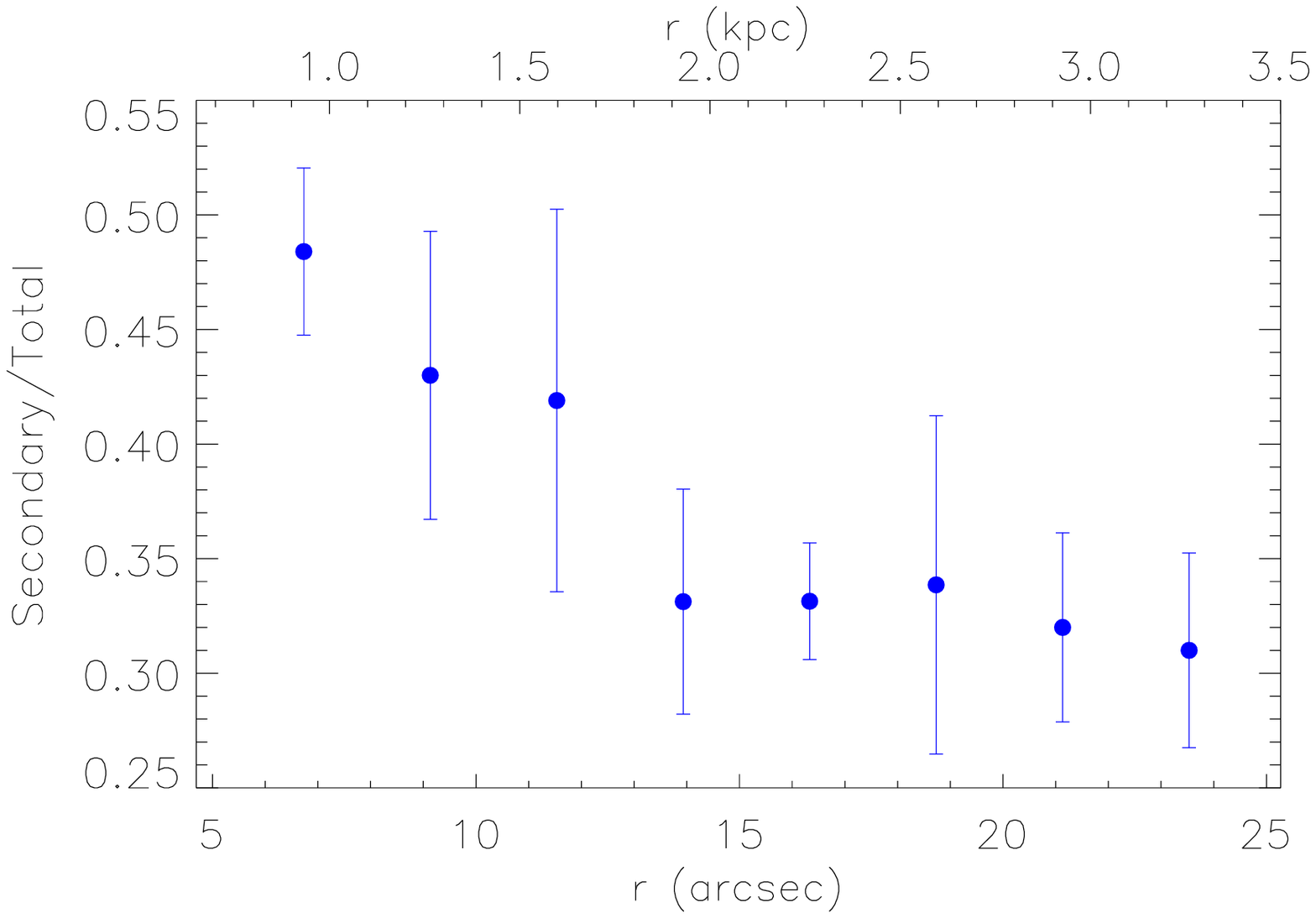}
\caption{Luminosity fraction of the secondary component over the total, as a function of 
radius.}
\label{fig:frac}
\end{figure}

\begin{figure}
\centering
\includegraphics[width=8.0cm]{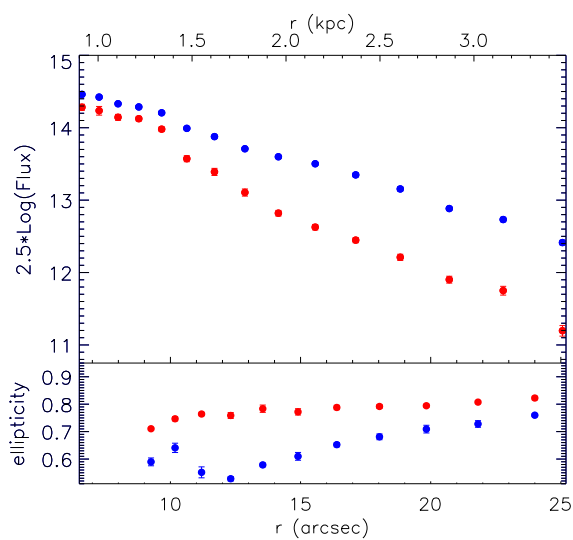}
\caption{Surface brightness (upper panel) and ellipticity (lower
  panel) radial profile of the primary (blue) and secondary (red)
  stellar components.}
\label{fig:radprof}
\end{figure}

\subsection{Stellar populations}
\label{sec:ssp}
\begin{figure*}
\centering
\includegraphics[width=8.0cm]{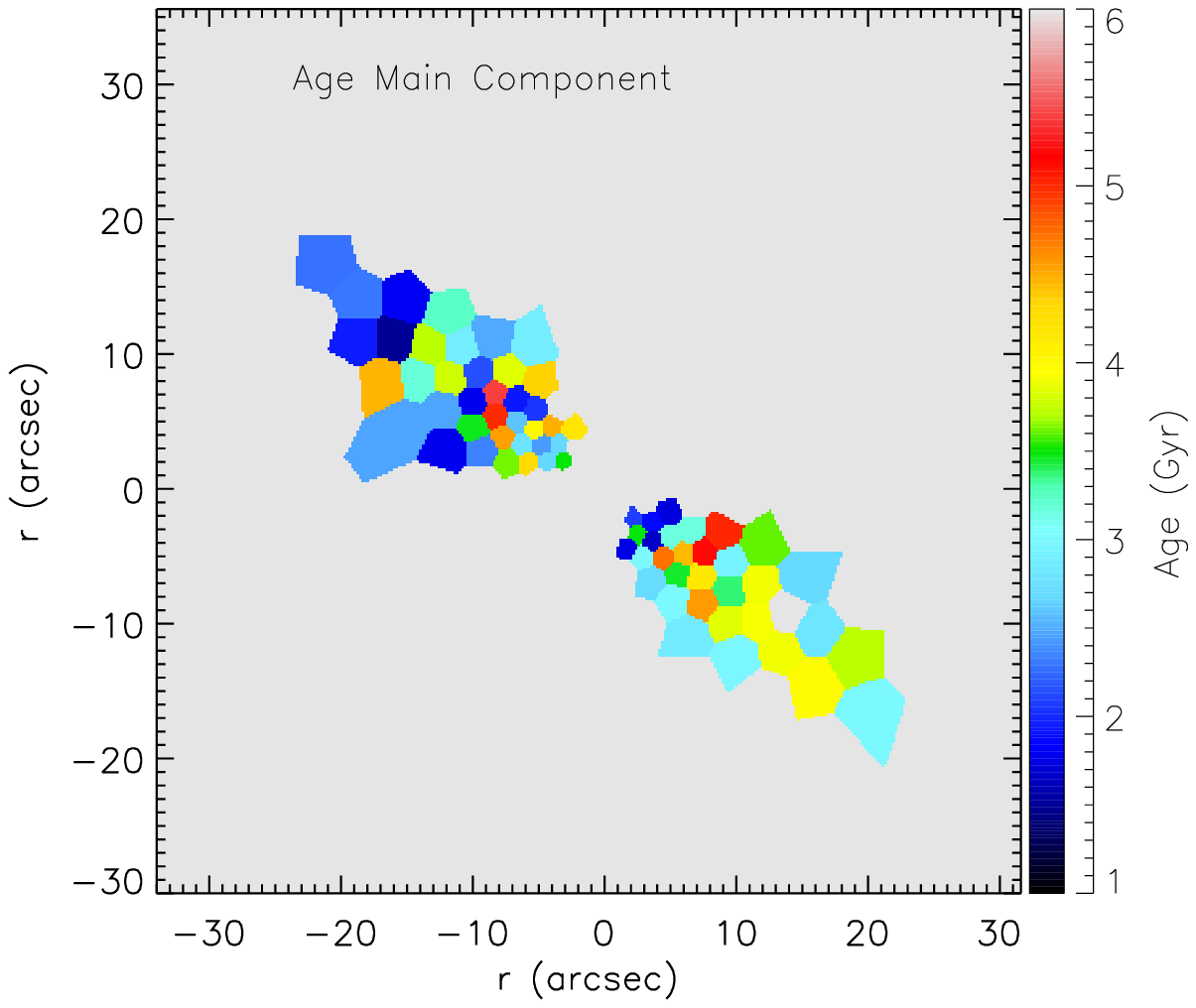}
\includegraphics[width=8.0cm]{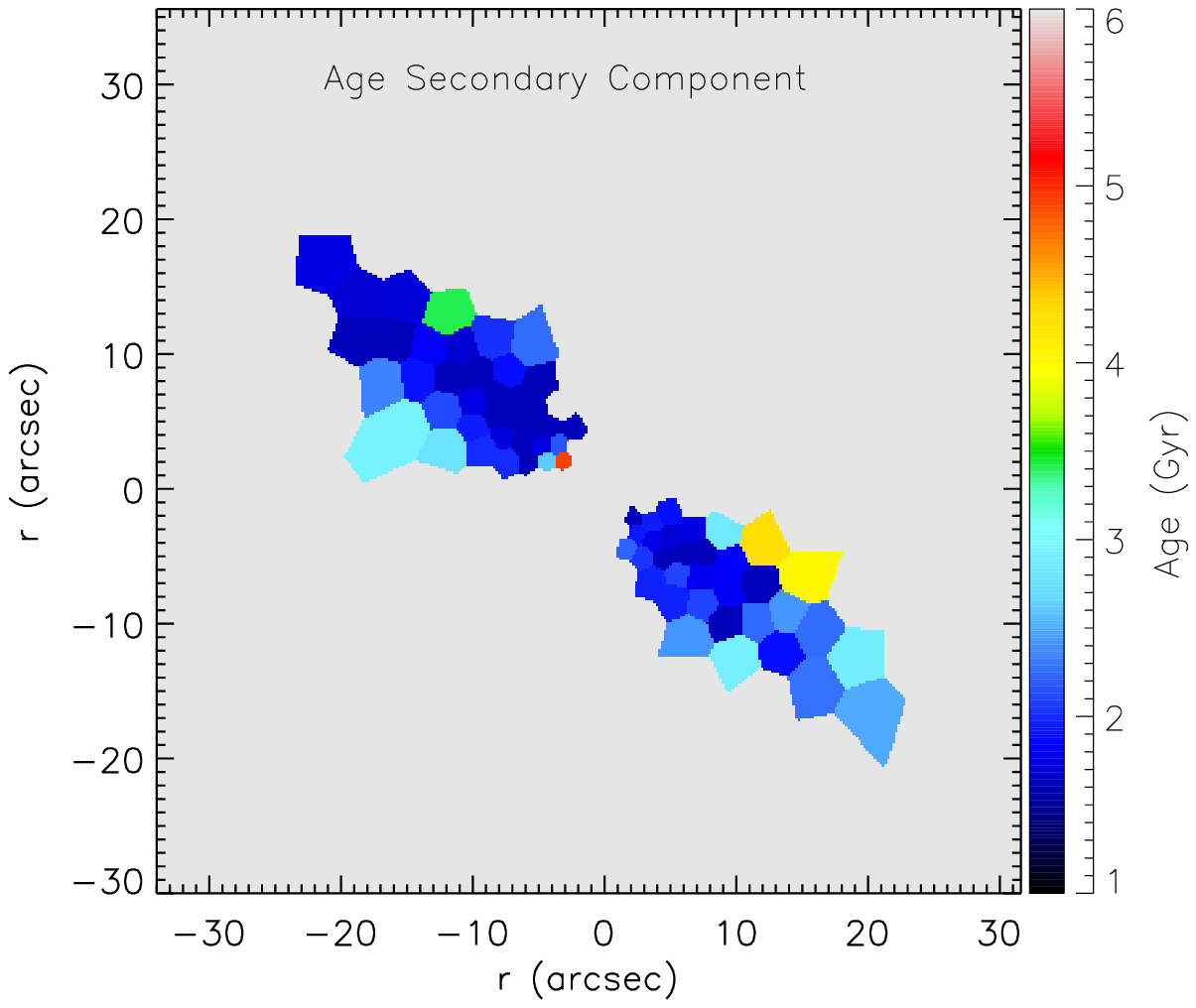}
\includegraphics[width=8.0cm]{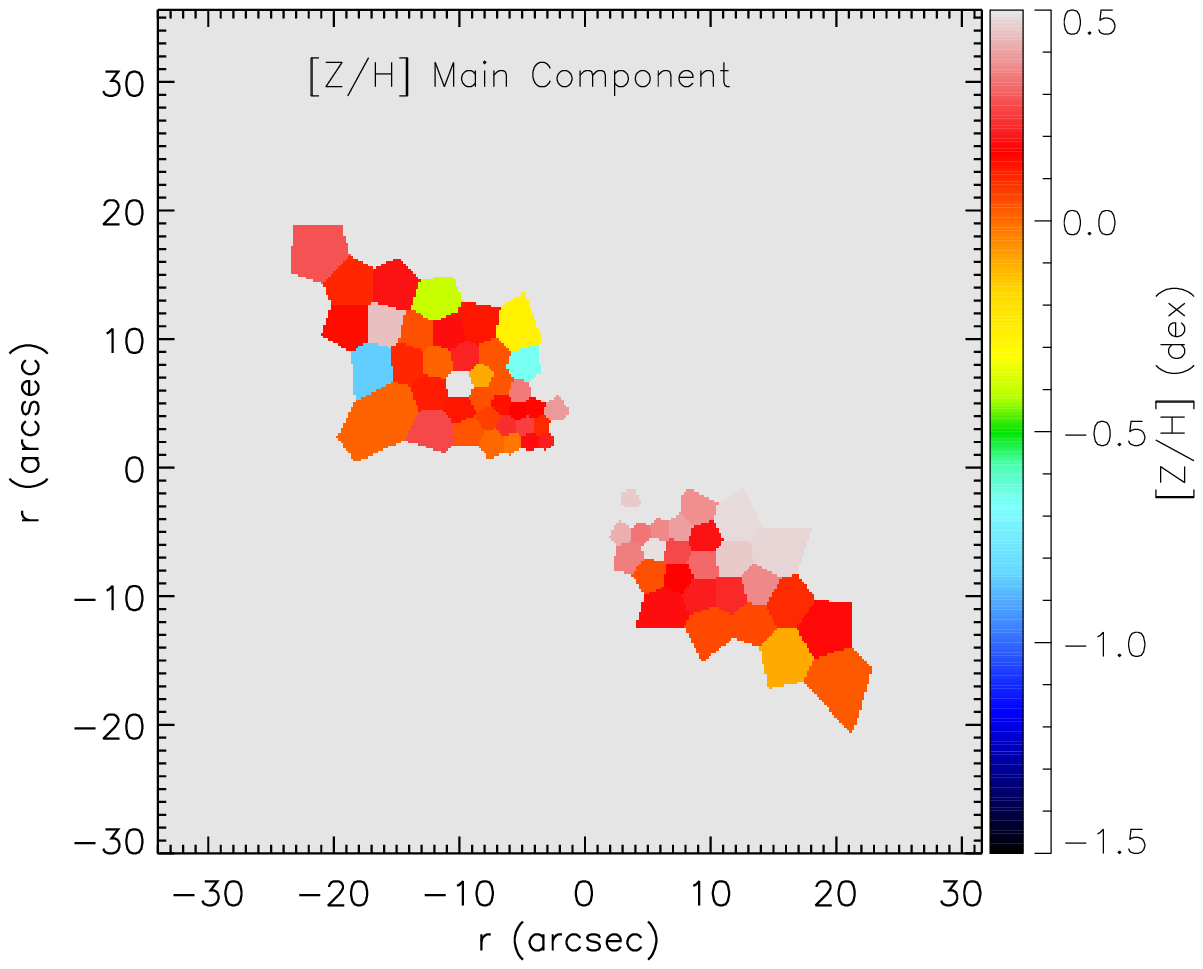}
\includegraphics[width=8.0cm]{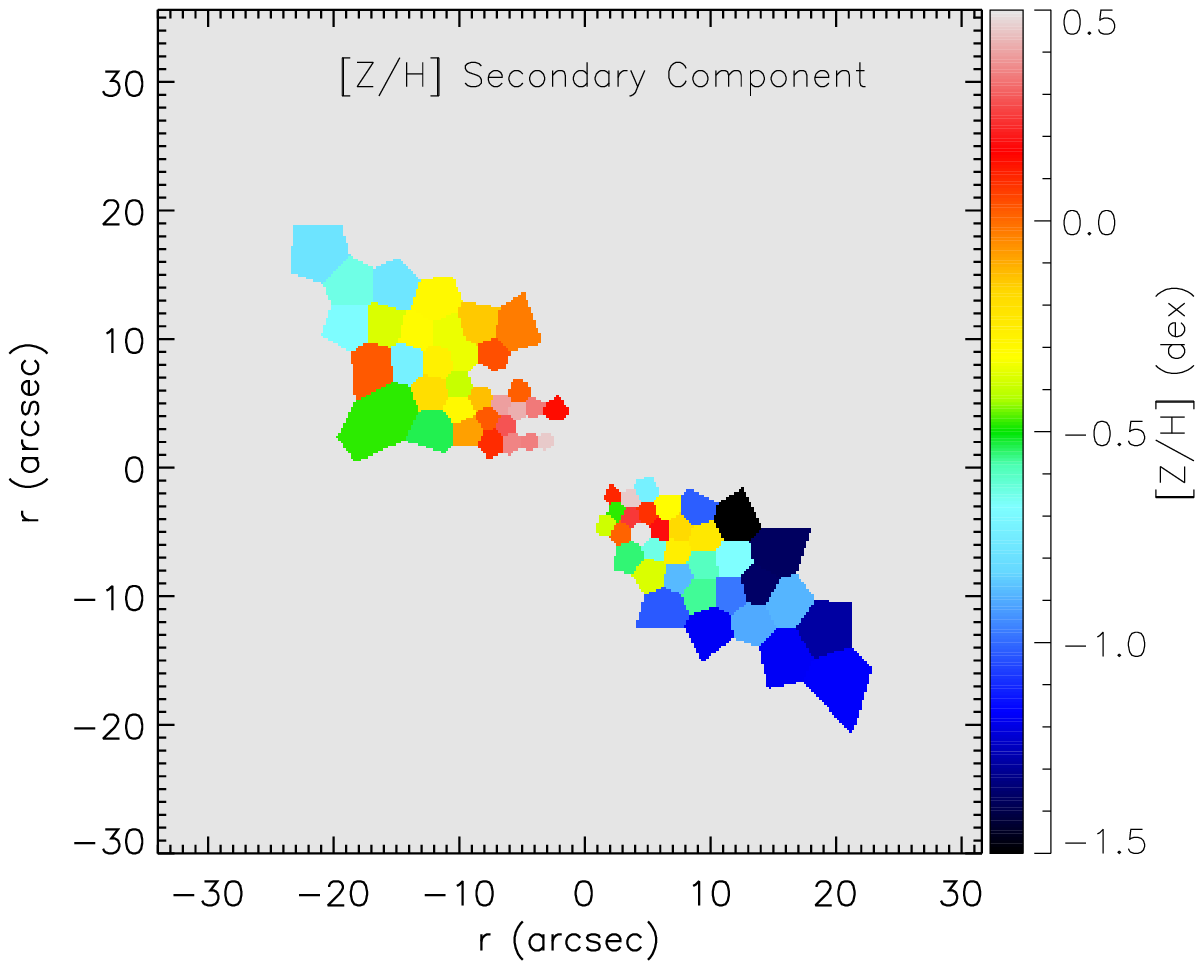}
\caption{Upper panels: Stellar population age for the primary (left) and secondary (right) components.
Lower panels: Stellar population metallicity for the primary (left) and secondary (right) components.}
\label{fig:agesmet}
\end{figure*}

For each bin we derived the map of the luminosity-weighted age and
metallicity of both the stellar components by comparing the
measurements of the line-strength indices \Hb, \Mgb, [MgFe]', and
$\langle$Fe$\rangle$ with the model predictions by \citet{Thomas+11}
for a single stellar population. In this parameter space the
  mean age and metallicity appear to be almost insensitive to the
  variations of the $\alpha/$Fe enhancement.  Age and metallicities
  were derived by linear interpolation between the model points using
  the iterative procedure described in \citet{2003A&A...407..423M}.
The final result is a 2D map of age and metallicity for
both the stellar components (Fig. \ref{fig:agesmet}).
\begin{figure}
\centering
\includegraphics[width=9.0cm]{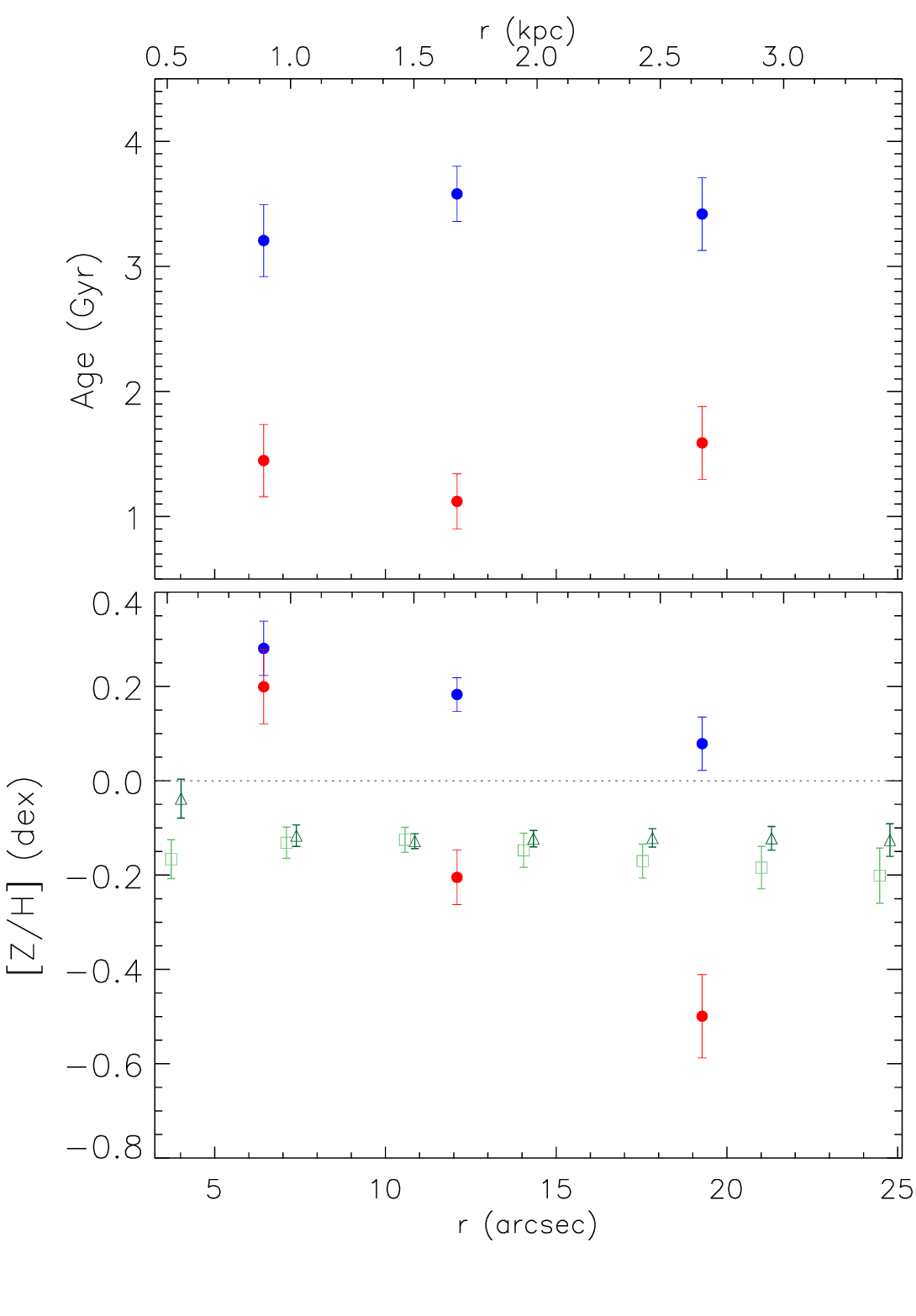}
\caption{Age (upper panel) and metallicity (lower panel) as a function
  of radius along the major axis of IC~719. Blue points represent the
  primary component and red points the secondary component. In the
  lower panel, open squares mark the gas-phase metallicity measured by
  means of the \niiig\ (green) and \niiig + \oiiig\ (light green) emission
  lines.}
\label{fig:popgradient}
\end{figure}

  The age and metallicity of both stellar components depend on the
  distance from the galaxy center. In order to analyze them we extracted their radial profiles
  from the 2D maps as follows.

For each
point in the field of view, we computed the deprojected distance from
the center of the galaxy assuming a disk major axis in the sky at PA=$53^\circ$,
 and the inclination of 77$^{\circ}$. We then
averaged age and metallicity values falling in the same deprojected
radial bins.  The results are shown in Fig. \ref{fig:popgradient}.

We find that the primary component has a luminosity-weighted age
 of about $3.5\,$Gyr constant with radius. The secondary
  component is also constant at about $1.5\,$Gyr.
The metallicity
of both stellar components decreases from a value of ([Fe/H]$\sim
0.2\,$dex) in the central region ($<1\,$kpc) to a more metal poor
value ([Fe/H]$\sim 0.1,-0.6\,$dex for the primary and secondary
components, respectively) at about $3\,$kpc .

  It is worth noting that there are some differences between our
  results and those presented by \citet{2013ApJ...769..105K}. Their
  long-slit data do not show a significant age difference of
    the two components in the radial range $5\arcsec < R < 25\arcsec$,
    whereas we observe different ages. Concerning the metallicity
    gradient, they observe a strong negative gradient for the primary
    component and we do not. The cause of this discrepancy is not
    clear. However, the large scatter in their measurements makes the
    comparison between the two analyses very difficult.

\subsection{Gas-phase metallicity}
\label{sec:gas_metal}
The gas-phase metallicity is one of the crucial observational
diagnostics of the current evolutionary state of galaxies; its
measurements can provide important information on the relationship
between the ionized gas component and the secondary stellar component
co-rotating with the gas.

To estimate the gas-phase metallicity, we used the empirical relations
between the intensity of the nebular lines and metallicity.  We
estimated the gas-phase metallicity following
\citet{2013A&A...559A.114M}, considering \niiip\ and \niiip +
\oiiig\ emission lines\footnote{
  $12+\log ($O/H$) =8.743+0.462\times \log ($\nii$/$\Ha$)$, and
$12+\log ($O/H$)=8.533-0.214\times \log ( ($\oiii$/$\Hb$) / ($Ha$/$\nii$) )$ }.
The gas-phase metallicity measured for IC~719 was converted to $[Z/{\rm H}]$
following \citet{sommariva2012}\footnote{$\log (Z/Z_\odot)=12+\log ({\rm O}/{\rm H}) - 8.69$}.
The $[Z/{\rm H}]_{gas}$ profile is
reported in Fig. \ref{fig:popgradient}. The gas metallicity obtained
from both diagnostics is consistently sub-solar ([Fe/H]$\sim
-0.15\,$dex) and constant with radius in agreement with \citet{2013ApJ...769..105K}.

\subsection{Star formation rate}
\label{sec:sfr}
The current SFR spatial distribution can show possible links between
the ionized gas and the stellar counter-rotating disk.  In addition,
the present-day SFR activity can provide clues about the epoch of gas
acquisition when considering scenarios in which the counter-rotating
disk formed {\it in situ} from acquired gas.  Finally, the SFR intensity
allows a comparison of IC~719 with normal (co-rotating) galaxies.

\begin{figure}
\includegraphics[width=9.0cm]{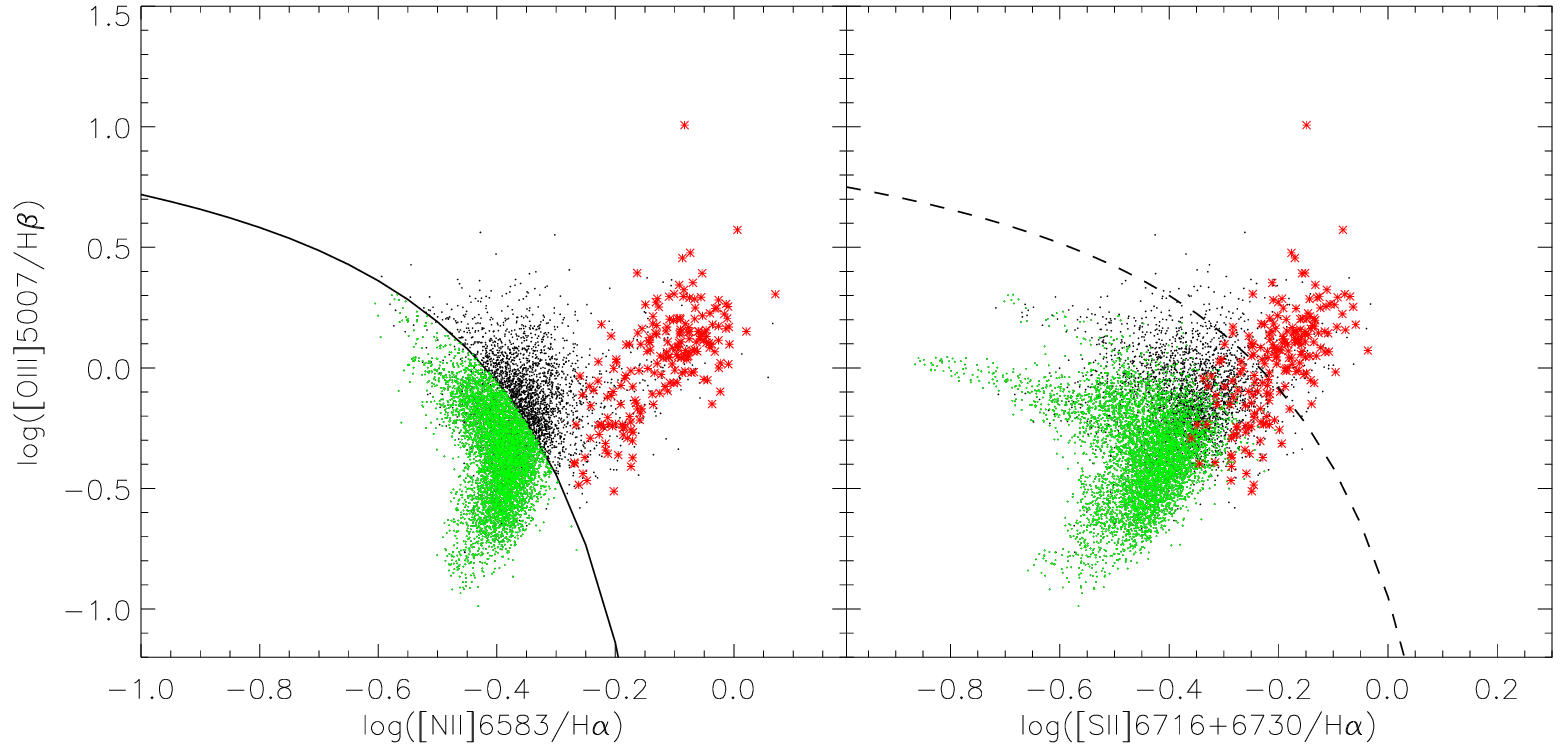}
\caption{BPT diagram on the \niiig , \oiiig, \Hb\ and \Ha\ lines. Red
  dots indicate the measurements belonging to the inner $3$\arcsec
  deprojected radius. The black full and dashed lines delimit the star-forming regions as indicated by \citet{2003MNRAS.341...54K} and
  \citet{2001ApJ...556..121K}. In both panels, green dots are the
  points that fall in the SFR region following both criteria.}
\label{fig:bpt}
\end{figure}

\begin{figure}
\includegraphics[width=9.0cm]{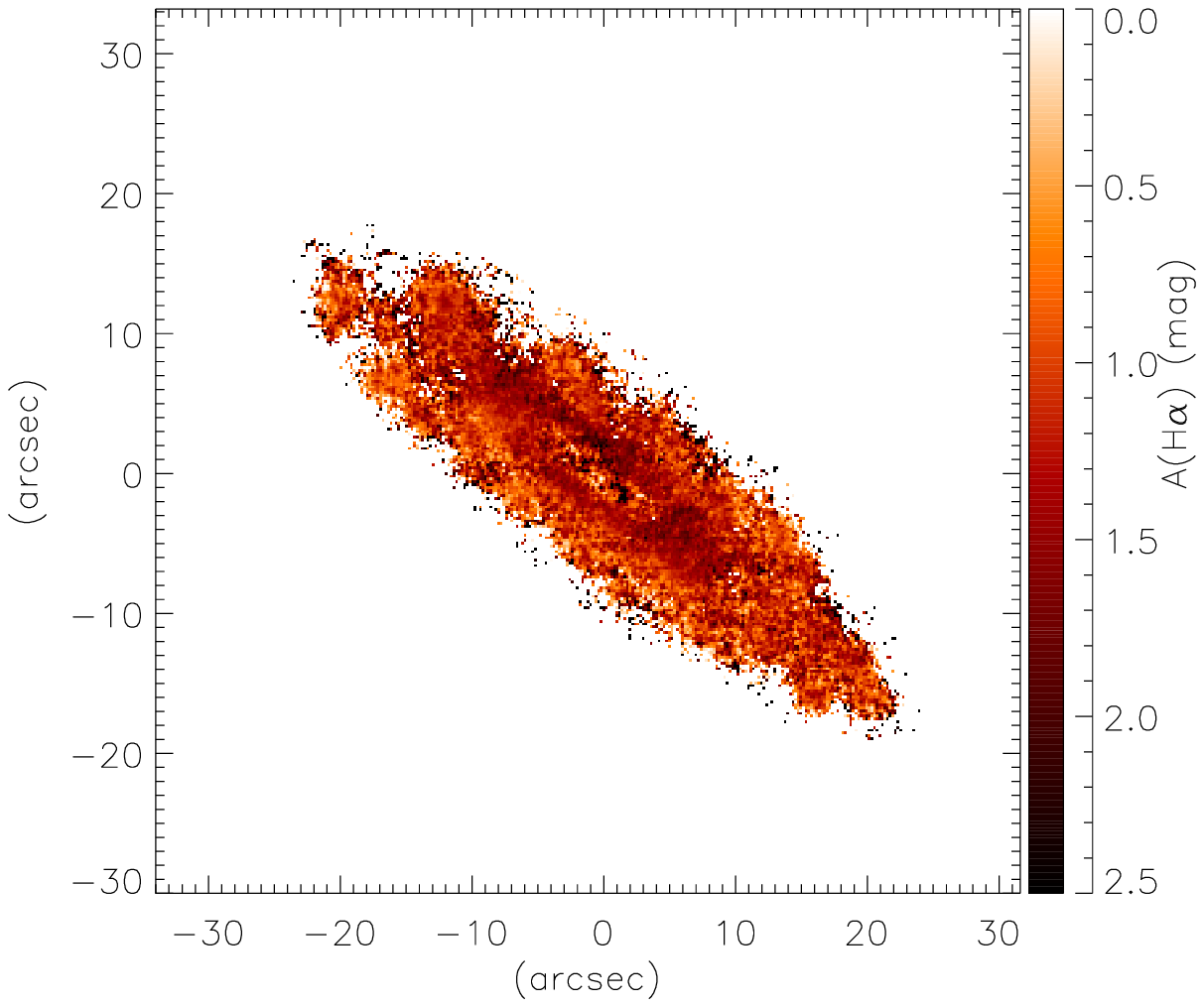}
\caption{Extinction map A(\Ha) derived following \citet{2015A&A...584A..87C}.}
\label{fig:est}
\end{figure}

\begin{figure}
\centering
\includegraphics[width=9.0cm]{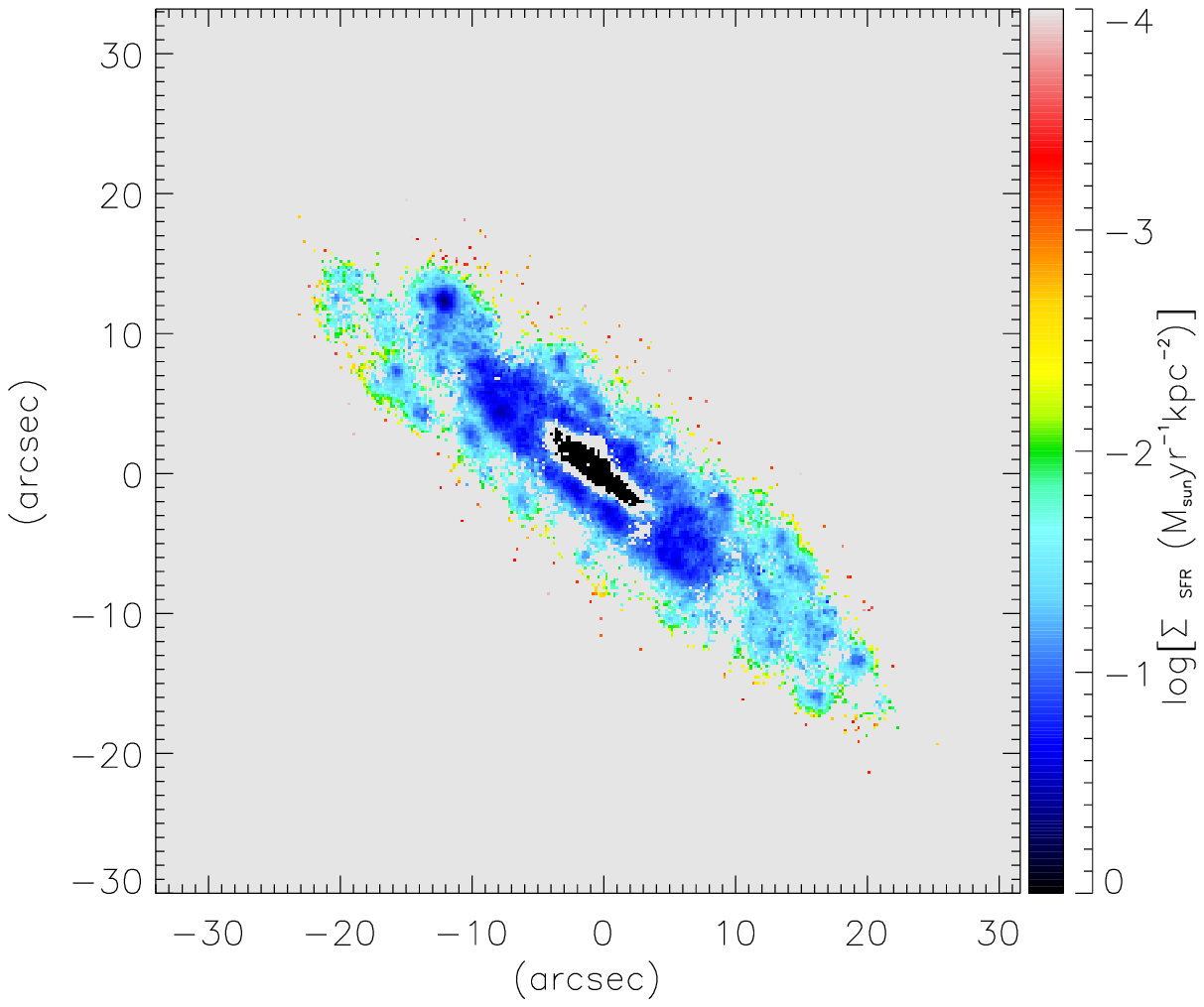}
\caption{Color-coded SFR surface density map of IC~719. Black dots indicate the region where the BPT diagnostic indicates
emission dominated by shocks. The gray region indicates where the signal was too low.  }
\label{fig:SFRmap}
\end{figure}

To derive the SFR in IC~719, we first identified the galaxy's star-forming region adopting the standard emission lines diagrams from
\citet[][BPT diagrams hereafter]{1981PASP...93....5B} as applied by
\citet{2003MNRAS.341...54K} and \citet{2001ApJ...556..121K}
(Fig.\ref{fig:bpt}). According to the BPT diagnostics, the regions
dominated by star formation describe a large elliptical annulus with an
inner semi-major axis of $5$\arcsec. We can trace star formation activity up to
the last measured point at about $30$\arcsec. Instead, the inner bulge
regions are dominated by shocks (Fig. \ref{fig:SFRmap}). The highest
star formation activity is located in a thin elliptical ring within
$\sim 7$ and $\sim 12$ \arcsec, with the same geometry and size as the
$K$-band ring observed by \citet{2013ApJ...769..105K}.

The SFR map was estimated considering the \Ha\ emission and
  Eq. 2 of \citet{1998ARA&A..36..189K} adopting the
extinction map obtained from the \Hb\ and \Ha\ ratio following
\citet{2015A&A...584A..87C} (Fig. \ref{fig:est}).  In
Fig. \ref{fig:SFRmap} we plot the resulting SFR map.

\begin{figure}
\centering
\includegraphics[width=8.0cm]{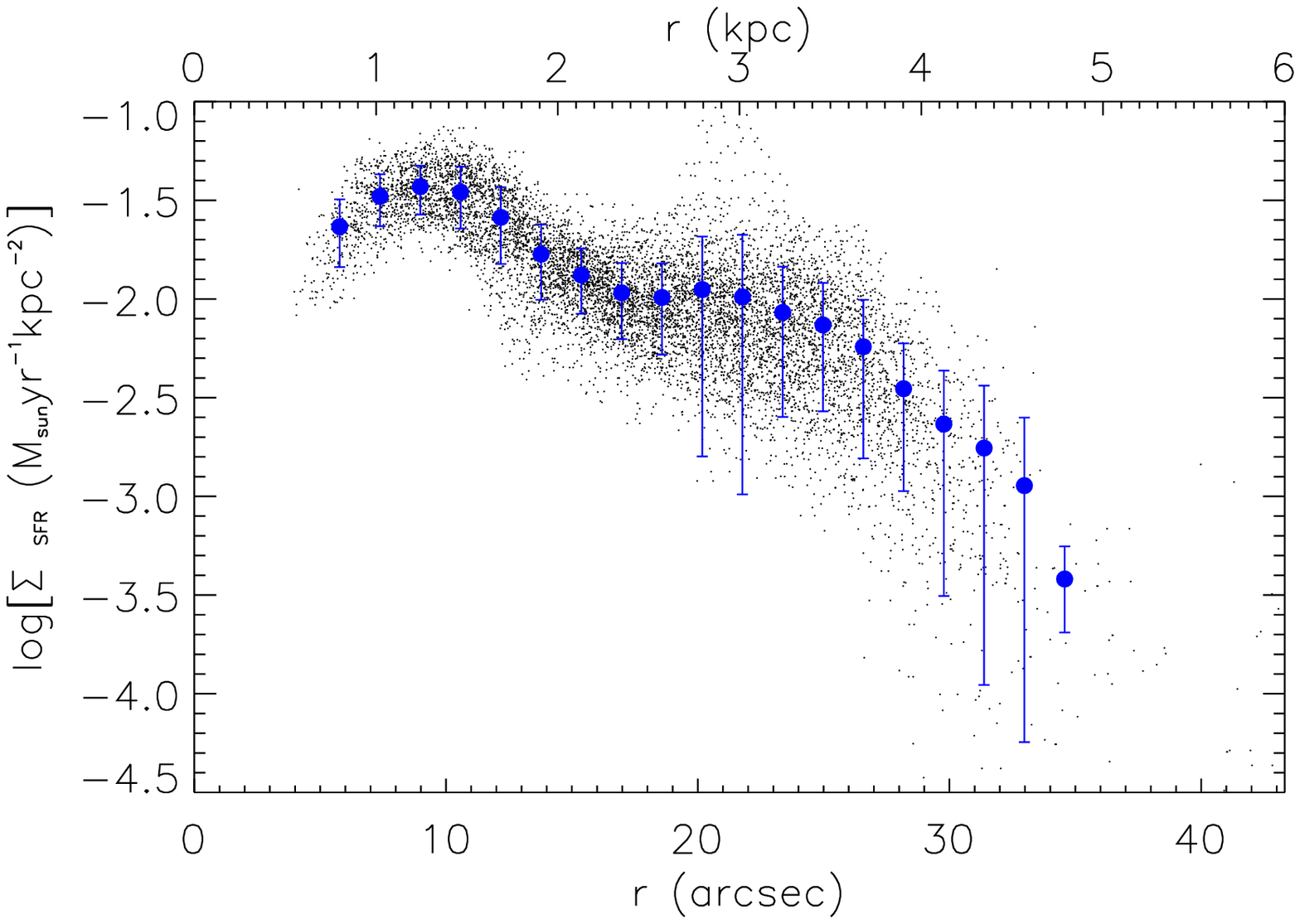}
\caption{Surface SFR as a function of the deprojected radius. Dots
   indicate measurements in each spaxel.  Blue dots and error-bars are
   the mean and the RMS computed in different radial bins. SFR has been
   deprojected for inclination.}
\label{fig:SFR}
\end{figure}

The total SFR of IC~719 is $\sim 0.5\,$M$_\odot$y$^{-1}$ , that is, an
  intermediate value between the star forming galaxies and ``retired''
  galaxies \citep{2016ApJ...821L..26C}.
The SFR surface density radial profile was derived in the
same way as the radial profiles of age and metallicity, and plotted in
Fig. \ref{fig:SFR}.  The maximum of the SFR is in the star-forming
ring; it has a value of $32\pm 8\,$M$_\odot $Gyr$^{-1}$pc$^{-2}$ at
$r=10$\arcsec\ and then decreases with a slope of about
$0.7\,$M$_\odot $Gyr$^{-1}$pc$^{-2}/$HLR (where HLR is the
half-light-radius that for IC~719 is $\sim1.9\,$kpc). These values and
trend are consistent with the values of a typical spiral galaxy
\citep{2015A&A...584A..87C}.

\section{Discussion and conclusion}
\label{sec:Discussion}

   We have analyzed the integral field data with a spectral
  decomposition technique in order to separate and independently
  analyze the properties of the two stellar components in the S0
  galaxy IC~719.

  Our analysis confirms the main results by
  \citet{2013ApJ...769..105K}: IC~719 has two extended stellar
  components that rotate along opposite directions. The ionized-gas
  co-rotates with the secondary stellar component, which is the
  younger, less luminous, and less massive. In addition, the analysis
  of the emission lines reveals that star formation is still active in
  IC~719 and it is the main source of ionization of the gas. All these
  elements strongly highlight the link between the secondary stellar
  component and the ionized-gas disk, supporting the idea that the
  secondary stellar component has been generated by the
    acquired gas disk, itself associated to a galaxy,
    in the case of a gas-rich minor merger, or not. In this latter case, we are
    not detecting the pre-merger stars of the dwarf galaxy.

  Thanks to the larger field of view of our MUSE data and the better
  spectral resolution in the CaT wavelength region used to
  derive the kinematics of the various components in IC~719, we are
  able to add some important elements with respect the previous
  studies on this galaxy.

  First, we measure a clear difference in the stellar velocity
  dispersion of the two stellar components. Specifically, the main component
  has larger velocity dispersion whereas the secondary component has
  smaller velocity dispersion.

  Second, our data allow us to study the 2D morphology of
  the stellar components, and therefore we have a complete description
  of their surface brightness profiles and geometry. The main
  component has an exponential profile with scale radius $r_H\sim
  1.5\,$kpc and an observed axial ratio in the range
  $q=0.40-0.30$. The secondary component has an exponential profile
  with scale $r_H\sim1.0\,$kpc and is thinner than the main
  component, having an observed axial ratio in the range
  $q=0.25-0.20$. This axial ratio is compatible with that of a very thin
  disk observed at an inclination of $i=77^{\circ}$.  In fact, if we
  use the value of $i=77^{\circ}$ to deproject the two disks, we find
  that the main component has an intrinsic axial ratio of 0.2-0.3
  while the secondary component has an axial ratio $\leq 0.15$.  While
  the main component has an intrinsic thickness comparable with a
  typical disk galaxy \citep{2013MNRAS.434.2153R}, the secondary
  component has an intrinsic thickness similar to that of a very thin
  disk, as we expect for ionized gas.
  
  The results on the stellar velocity dispersion and flattening put
  some important constraints on the formation scenario and strengthen
  previous conclusions. In the binary galaxy merger scenario
  \citep{2009MNRAS.393.1255C}, the ionized gas component is expected to
  co-rotate with the thicker, dynamically hotter, and more massive
  component.  This is true for a perfectly co-planar geometry of the
    encounter. \citet{2011A&A...535A...5Q} studied the case of 1:10
    minor mergers with different geometries.  In general, they find that
    the acquired counter-rotating stars form a counter-rotating thick stellar disk.
  Our measurements of the velocity dispersion and flattening
  indicate the opposite and are therefore consistent with the scenario
  in which the counter-rotating stars formed from gas
  acquired from either a gas cloud or a gas-rich dwarf galaxy.
  
Star formation and gas accretion are still ongoing in IC~719. In fact,
as pointed out by \citet{2013ApJ...769..105K}, the galaxy is
surrounded by a considerable amount of HI \citep{2009A&A...498..407G}
that is probably the source of the ionized gas. It seems to form a
common envelope with the HI surrounding the companion IC~718 which is
located at a projected distance of $96\,$kpc.

 In their work, \citet{2013ApJ...769..105K} advocate the scenario in
 which the external gas was accreted by IC~719 in two distinct
 episodes, both followed by star formation, in contrast with the
 alternative scenario of a single accretion episode followed by two
 bursts of star formation. In this picture, the HI is likely still in
 the process of falling onto the galaxy.  The main argument in favor
 of two accretion events is that the metallicity of the ionized gas is
 lower than the metallicity of the secondary stellar component; it
 would be the opposite if the generation of stars were formed by a
 second burst using gas processed by the first generation of stars.
 In this picture, the different radial behavior of the
   metallicity of the gas-phase with respect to the secondary stellar
   component may indicate that the gas experienced, after the star
   formation, prolonged or continuous gas acquisition (like in
   NGC~5719) and/or enrichment caused by the stellar evolution.

  The age gradient we measure is consistent with the scenario proposed
  by \citet{2013ApJ...769..105K}. The accretion and star formation
  occurred $\sim 4\,$Gyr ago and created the first generation of stars in
  the secondary component. The second accretion and burst dominated
  the inner $\sim 2.5\,$kpc, where younger luminosity-weighted ages are
  measured and the star formation activity is higher. The fact that
  the stellar metallicity is higher than that of the gas phase in the
  inner $\sim 2\,$kpc is also consistent with this picture: the stars
  in the central region originated from the accreted gas and gas
  that has been re-processed and enriched by the previous generation
  of stars, resulting in a higher metallicity. The flat radial
  profile of the gas-phase metallicity supports the idea that the two
  accreted gas masses come from the same source, most likely the HI
  cloud surrounding IC~719.

  It is interesting to study how IC~719 compares with other galaxies that host
  counter-rotating stellar components.  We found a similar system in
  NGC~5719 where an HI bridge is connecting it to NGC~5718. NGC~5719
  hosts a counter-rotating disk younger than the main galaxy and with
  a ring of star formation of about $1\,$kpc radius. A third galaxy that
  shows similar behavior is NGC~3595. Here the star-forming ring has
  a diameter of about $1\,$kpc.  \citet{2009A&A...498..407G} find this
  galaxy to be an early-type
  galaxy given the great amount of HI contained within. \citet{2014Ap&SS.354...83B} considered this latter object as a
  case of rejuvenation induced by the acquisition of gas.  In this
  respect, it would be interesting to know how IC~719 would look like
  if the gas was acquired in pro-grade orbits. Most likely it would appear like a
  normal Sb galaxy or possibly host a
  star-forming nuclear ring \citep[see][and references
    therein]{2017arXiv170501251M}.  For instance,
  \citet{2015A&A...575A..16M} studied the formation of gas rings in an S0
  galaxy as a consequence of a gas-rich minor merger for both
  prograde and retrograde encounters. They found, by means of
  N-body/smoothed particle hydrodynamics simulations, that in both
  cases a warm gas ring forms and causes a rejuvenation of the stellar
  population. 
For NGC~448, \citet{2016MNRAS.461.2068K}
suggested a different formation mechanism. In this case, they exclude a cosmic
gas filament as the origin of the acquired gas. They preferred a satellite
merger with slow gas accretion as the origin of the counter-rotating
stellar disk. On the other hand, NGC~448 shows no star formation
activity and an old stellar population ($9\,$Gyr) for both components and
can be seen as the evolution of systems like IC~719. This is probably
true for other counter-rotating systems such as NGC~4550 and NGC~4191 that
show a small amount of gas, no ongoing star formation, and a relatively
old stellar population in the acquired component.

 In summary, IC~719 represents a clear case of an S0 galaxy hosting two
 stellar counter-rotating disks. Thanks to the large field of view and
 wavelength coverage of the MUSE spectrograph, we are able to add
 further constraints to the formation mechanisms. In particular, we can
 unveil the structural and kinematical differences between the two
 stellar components. We confirm the previous scenario in which the
 secondary stellar component of IC~719 formed via multiple accretion
 of gas coming from the surrounding HI cloud followed by star
 formation episodes.

\begin{acknowledgements}
  This work was supported by Padua University through grants
  60A02-5857/13, 60A02-5833/14, 60A02-4434/15, and CPDA133894.  AP and
  LC thank the European Southern Observatory and Department of Physics
  and Astronomy of the University of Padova, respectively, for
  hospitality during the development of this paper.  LM and EMC
  acknowledge financial support from Padua University grants CPS0204
  and BIRD164402/16, respectively.
  
\end{acknowledgements}

\bibliographystyle{aa} 
\bibliography{ic719}

\end{document}